\documentclass[secnumarabic,amssymb,nature, nobibnotes,aps,preprint,letterpaper,preprintnumbers,amsmath,superscriptaddress,showpacs,floatfix]{revtex4-1}
\usepackage{amssymb}
\usepackage{amsmath}
\usepackage{graphicx}
\usepackage{dcolumn}
\usepackage{bm}
\usepackage{multirow}
\usepackage{epsfig}
\usepackage{xcolor}
\usepackage{blkarray}
\newsavebox{\overlongequation}

\renewcommand{\thetable}{\arabic{table}}

\begin{document}

\title[Chemical bonding and Born charge in 1T-HfS$_2$]{Chemical bonding and Born charge in 1T-HfS$_2$}

\author{S. N. Neal}
\affiliation{Department of Chemistry, University of Tennessee, Knoxville, Tennessee, USA}

\author{S. Li}
\affiliation{Department of Chemical Engineering and Materials Science, University of Minnesota, Minneapolis, Minnesota, USA}

\author{T. Birol}
\affiliation{Department of Chemical Engineering and Materials Science, University of Minnesota, Minneapolis, Minnesota, USA}

\author{J. L. Musfeldt}

\affiliation{Department of Chemistry, University of Tennessee, Knoxville, Tennessee, USA}
\affiliation{Department of Physics and Astronomy, University of Tennessee, Knoxville, Tennessee, USA}
\email{musfeldt@utk.edu}

\date{\today}



\begin{abstract}

We combine infrared absorption and Raman scattering spectroscopies
to explore the properties of the heavy transition metal dichalcogenide 1T-HfS$_2$. 
We employ the LO-TO splitting of the $E_u$ 
vibrational mode along with a reevaluation of mode mass, unit cell volume, and dielectric constant to reveal the Born effective charge. We find $Z^*_{\rm{B}}$ = 5.3$e$, in excellent agreement with complementary first principles calculations. In addition to resolving controversy over the nature of chemical bonding in this system, we decompose Born charge into polarizability and local 
charge. We find $\alpha$ = 5.07 \AA$^3$ and $Z^{*}$ = 5.2$e$, respectively. Polar displacement-induced charge transfer from sulfur $p$ to hafnium $d$ is responsible for the enhanced  Born charge compared to the nominal 4+ in hafnium. 1T-HfS$_2$ is thus an ionic crystal with strong and dynamic covalent effects. Taken together, our work places the vibrational properties of  1T-HfS$_2$ on a firm foundation and opens the door to understanding the properties of tubes and sheets.

\vspace{.25in}

\end{abstract}


\maketitle \clearpage

\section*{Introduction}

While 3$d$ transition metal oxides and chalcogenides display strong electronic correlations, narrow band widths, and robust magnetism, 4 and 5$d$ systems are recognized for strong spin–orbit coupling, increased hybridization, and more diffuse orbitals. As a result, materials that contain 4- and 5$d$ centers often have enhanced or emergent properties \cite{Kim2008,Birol2015,Zwartsenberg2020,Cao2018, Cao2000, Chen2020, Singleton2016,ONeal2019, Ma2018}. Transition metal dichalcogenides such as MoTe$_2$, IrTe$_2$, HfSe$_2$, and PtSe$_2$  are  of great interest for their unconventional chemical bonding and hybridization, topology, 
multiferroicity, and tendency toward complex dimerization patterns \cite{Yao2018,Zhu2017, ChenChen2017, Ideta2018, Pascut2014,Jiang2020, Heine2020, Koz2020}. Within this class of materials, 1T-HfS$_2$ has attracted particular attention as an analog of HfO$_2$ - a highly polarizable gate dielectric
\cite{Cheema2020, HfO2020}.

1T-HfS$_2$ is a layered material with a  $P{\bar 3}m1$ (\#164) space group at 300 K \cite{Iwasaki1982}. Each Hf$^{4+}$ ion has $D_{3d}$ site symmetry and is located at the center of a S$^{2-}$  octahedron. 
The van der Waals gap is 3.69 \AA, and the sheet thickness is 2.89 \AA.
Photoemission studies reveal an indirect band gap of 2.85 eV between $\Gamma$ and M/L, which varies slightly from the $\approx$ 2 $eV$ optical gap  \cite{traving2001combined}. 
1T-HfS$ _2$ forms a high performance transistor with excellent current saturation \cite{Kanazawa2016}. The carrier mobility is on the order of 1800 cm$^2$V$^{-1}$s$^{-1}$ - much higher than MoS$_2$ and thickness dependent as well \cite{Xu2015,Najmaei2018}. 
Group theory predicts that  at the $\Gamma$ point 1T-HfS$_2$ has vibrational modes with symmetries of $A_{1g}$ + $E_g$ + $A_{2u}$ + $E_u$. The $A_{1g}$ + $E_g$ modes are Raman-active, and the $A_{2u}$ + $E_u$ modes are infrared-active \cite{Iwasaki1982,Roubi1988}.
Despite many years of work, there are a surprising number of unresolved questions about 1T-HfS$_2$ 
- even in single crystal form. In the field of vibrational spectroscopy,  there is controversy about mode assignments, the role of resonance in creating hybrid modes, the presence or absence of surface phonons, and the use of this data to reveal the Born effective charge ($Z_B^*$). 
As an example, Born effective charges between 3.46$e$ and 5.5$e$ have been reported by various experimental \cite{Uchida1978, Lucovsky1973} and theoretical \cite{Chen2016} groups. 
Evidence for the degree of ionicity (or covalency) is both interesting and important because 5$d$ orbitals tend to be more diffuse than those of their 3$d$ counterparts. 
Within this picture, 1T-HfS$_2$ has the potential to sport significant covalency.
High pressure Raman scattering spectroscopy reveals a first-order phase transition near 11 GPa and different ${\partial}{\omega}/{\partial}P$'s (and thus mode Gruneisen parameters) for the hybrid $E_u$ and fundamental $A_{1g}$ modes \cite{Ibanez2018}.  At the same time, variable temperature Raman scattering spectroscopy  shows a systematic blueshift of the spectral features down to 100 K, except for the large $A_{1g}$ mode near 330 cm$^{-1}$  which redshifts \cite{Najmaei2018, Ibanez2018}. 
In few- and single-layer form, 1T-HfS$_2$ is suitable for high-performance transistors \cite{Kanazawa2016, Fu2017, Zhang2019}, displays a direct gap (rather than indirect as in the bulk) \cite{Kang2015}, exhibits photocatalytic behavior appropriate for water splitting \cite{Singh2016}, reveals applications in photodetection \cite{photodetectorHf2018}, is susceptible to strain effects \cite{Wu2017}, and is useful in N, C, and P surface adsorption \cite{Berwanger2020}. This system can be integrated into van der Waals heterostructures and grown vertically as well \cite{Zheng2016, Zhang2019, Lei2019}.

In order to explore the vibrational properties of 1T-HfS$_2$, we measured the infrared absorption and Raman scattering response  
and employed the results to evaluate Born effective charge. We find that $Z^*_{\rm{B}}$ = 5.3$e$, in excellent agreement with our complementary first principles calculations. In order to understand how $Z^*_{\rm{B}}$ relates to the nominal 4+ charge of the Hf center, we employ a Wannier function analysis to  project out the different orbital  contributions. This analysis reveals that the sulfur $p$ orbital transfers charge to the cation and that this contribution enhances the Born charge beyond the nominal value. Decomposing  $Z^*_{\rm{B}}$ into polarizability and local charge, we find that there is strong ionicity as well as significant covalency in this system. Both are quite different than in 2H-MoS$_2$ - probably on account of spin-orbit coupling. We also identify two weak structural distortions near 210 and 60 K evidenced by subtle frequency shifts of the $E_g$ and $A_{1g}$ vibrational modes as well as changes in the phonon lifetimes. Taken together, these findings resolve controversy over the nature of chemical bonding in 1T-HfS$_2$ and clarify the role of the 5$d$  center in the process.

\section*{Results and Discussion}

\subsection*{First-principles predictions of charge and bonding}

\begin{figure*}[tbh]
\centering
\includegraphics[width = 5.5in]{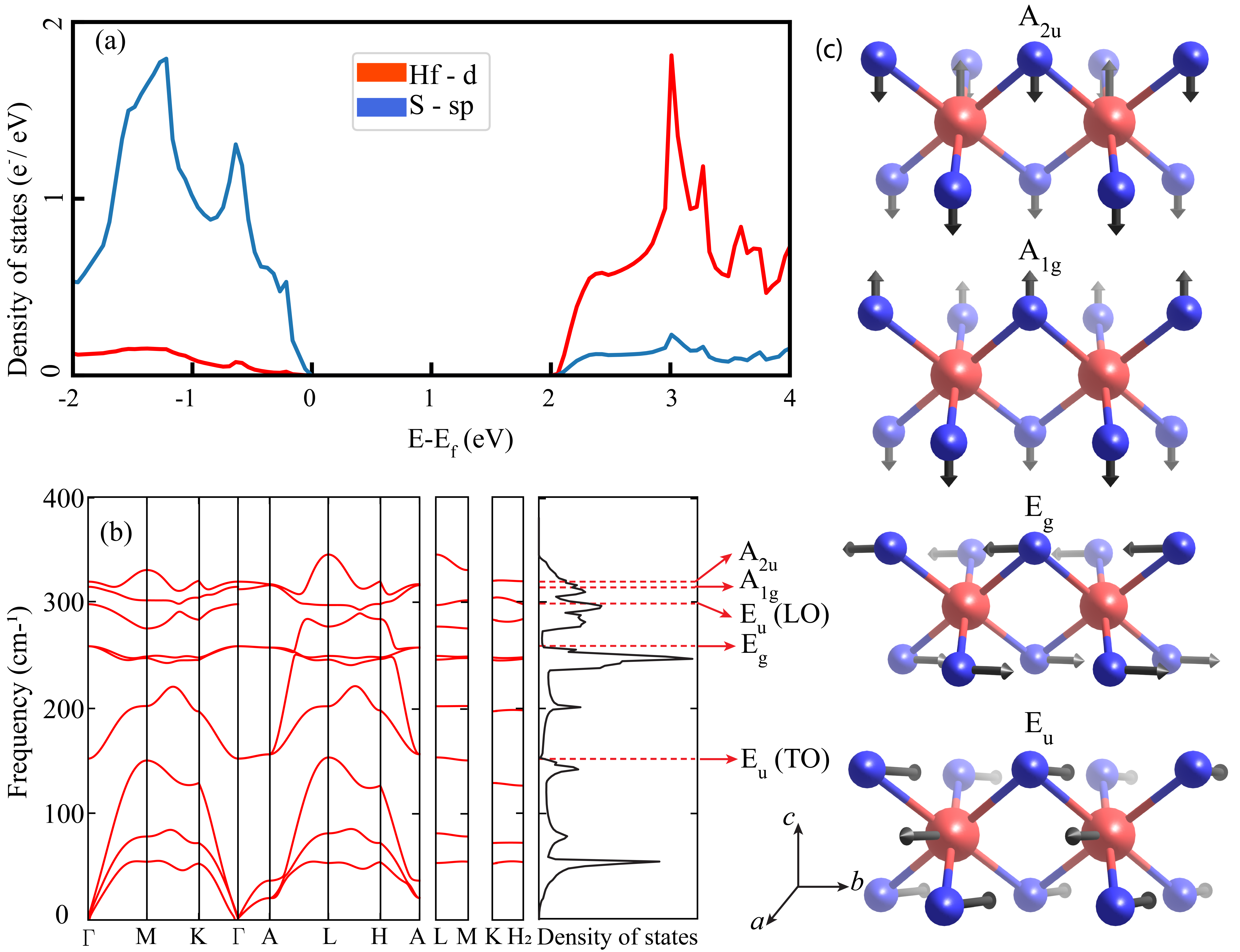}
\caption{{\bf{\label{fig:dft} Density of states and displacement patterns of 1T-HfS$_2$.}} (a) The electronic density of states of 1T-HfS$_2$. Different colors indicate different orbitals of the two types of atoms. (b) The phonon spectrum of 1T-HfS$_2$ and its density of states. The $\Gamma$-point phonon mode frequencies are marked with dashed lines. (c) Displacement patterns of several $\Gamma$-point phonon modes. Blue atoms are S and the red atoms are Hf. The arrows are not scaled to real displacement amplitudes. Numerical details are available in Supplementary Information. 
}
\end{figure*}

Figure~\ref{fig:dft}(a) displays the projected density of states of 1T-HfS$_2$ computed using density functional theory and atom centered local projectors \cite{Kresse1999}. The bands can be assigned Hf and S character easily, and the degree of hybridization between the atoms is not dominant (albeit nonzero). This reveals the strong ionic nature of this system: The valence band is composed mainly of S-$p$ orbitals whereas the conduction band is predominantly Hf-$d$ in character. We find a band gap of 2.05~eV using the HSE06 functional. This is  consistent with the small electronegativity of the Hf (1.3 on the Pauling scale) compared to that of S (2.6 on the Pauling scale). 

Despite the apparent ionicity of the density of states, the dynamical Born effective charges in 1T-HfS$_2$ are anomalous.  We find an in-plane value for  the Hf ions as $Z^*_{B,xx} = +6.4e$. By contrast, the out-of-plane value for Hf $Z^*_{B,zz}$ is only 2.0$e$. This reveals that either (i) the Hf ions are strongly polarizable, or (ii) the small degree of covalency is strongly dependent on ionic displacements \cite{ghosez1998}.

Figure~\ref{fig:dft}(b) displays the phonon dispersions of 1T-HfS$_2$. 
While the spectrum is highly dispersive within the plane (for instance, in the $\Gamma$-M direction) it is much less so in the out-of-plane direction (for instance, along $\Gamma$-A). This difference is a natural consequence of the layered crystal structure and is the origin of the spikes in the phonon density of states [right panel, Fig.~\ref{fig:dft}(b)]. One aspect of these predictions that will be important for later discussion is the mode order around the $A_{1g}$ fundamental. Notice that the $E_u$ feature is predicted to be below the $A_{1g}$ mode, whereas the $A_{2u}$ mode is predicted to be above the $A_{1g}$ fundamental. These features are labeled in Fig.~\ref{fig:dft}(b).
Theoretical phonon frequencies are in excellent agreement with our measured results [Table~\ref{Table 1}].

Lattice dynamics can be used to gain information about the chemical bonding in crystals. In 1T-HfS$_2$, the $E_u$ optical mode is due to the in-plane vibrations of the Hf cations  against the S anions [Fig.~\ref{fig:dft}(c)]. The frequency difference between the longitudinal and transverse optical modes (the LO-TO splitting) depends on the permittivity, as well as the dynamical charges of the ions. Formally, the LO-TO splitting stems from a non-analytic term added to the dynamical matrices of ionic insulators at the zone center \cite{baroni2001phonons}:
\begin{equation}
	D^{nan}_{st, \alpha\beta} = \lim_{q\to 0} \frac{1}{\sqrt{m_s m_t}}\frac{4\pi}{\Omega}e^2\mathbf{\frac{(q\cdot Z^*_{B,s})_\alpha(q\cdot Z^*_{B,t})_\beta}{q\cdot\varepsilon(\infty)\cdot q}}
\end{equation}
Here the $s, t$ are atomic indices and $\alpha, \beta$ are cartesian directions. $\mathbf{Z_{B}^*}$ is the Born effective charge and $\mathbf{q}$ is the wavevector. With this extra term present in the dynamical matrix, the two-fold degeneracy of the $E_u$ optical modes is lifted:

\begin{equation}
\omega_{Eu,TO}^2 - \omega_{Eu,LO}^2 = D_{Eu}^{nan} = \frac{4\pi}{\Omega \varepsilon_{xx}(\infty)}e^2 \left(\sum_i \frac{\mathbf{u_s\cdot Z^*_{B,i}}}{\sqrt{m_i}}\right)^{2},
\label{eq:LO-TO}
\end{equation}
where $u_i$ is the displacement vector of $E_u(TO)$ mode. By considering the fact that $Z_{Hf,xx}^*$ is always equal to $2\cdot Z_{S,xx}^*$ because of charge neutrality, and the symmetry imposed form of the $E_u$ displacement pattern, we can write this equation for $\mathbf{q}\parallel \hat{x}$ as:

\begin{equation}
\omega_{TO}^2 - \omega_{LO}^2 = \frac{4\pi}{\Omega \varepsilon_{xx}(\infty)}e^2 \frac{Z_{Hf^*,x}^2}{m^*},
\label{eq:LO-TO2}
\end{equation}

\noindent 
where the $m^*$ is the effective mass, determined by:

\begin{equation}
    \frac{1}{m^*} =  \frac{1}{m_{Hf}}+\frac{1}{2m_S}.
    \label{Mass}
\end{equation}

\noindent 
In this case, the effective mass is $47.02~u$. A general expression for effective mass is provided in Supplementary Information. We use this result along with  Equation~\ref{eq:LO-TO2} to analyze the  experimental Born effective charge below. Gaussian units are employed.


\subsection*{Infrared properties of 1T-HfS$_2$}

\begin{figure*}[tbh]
\centering
\includegraphics[width = 6.0in]{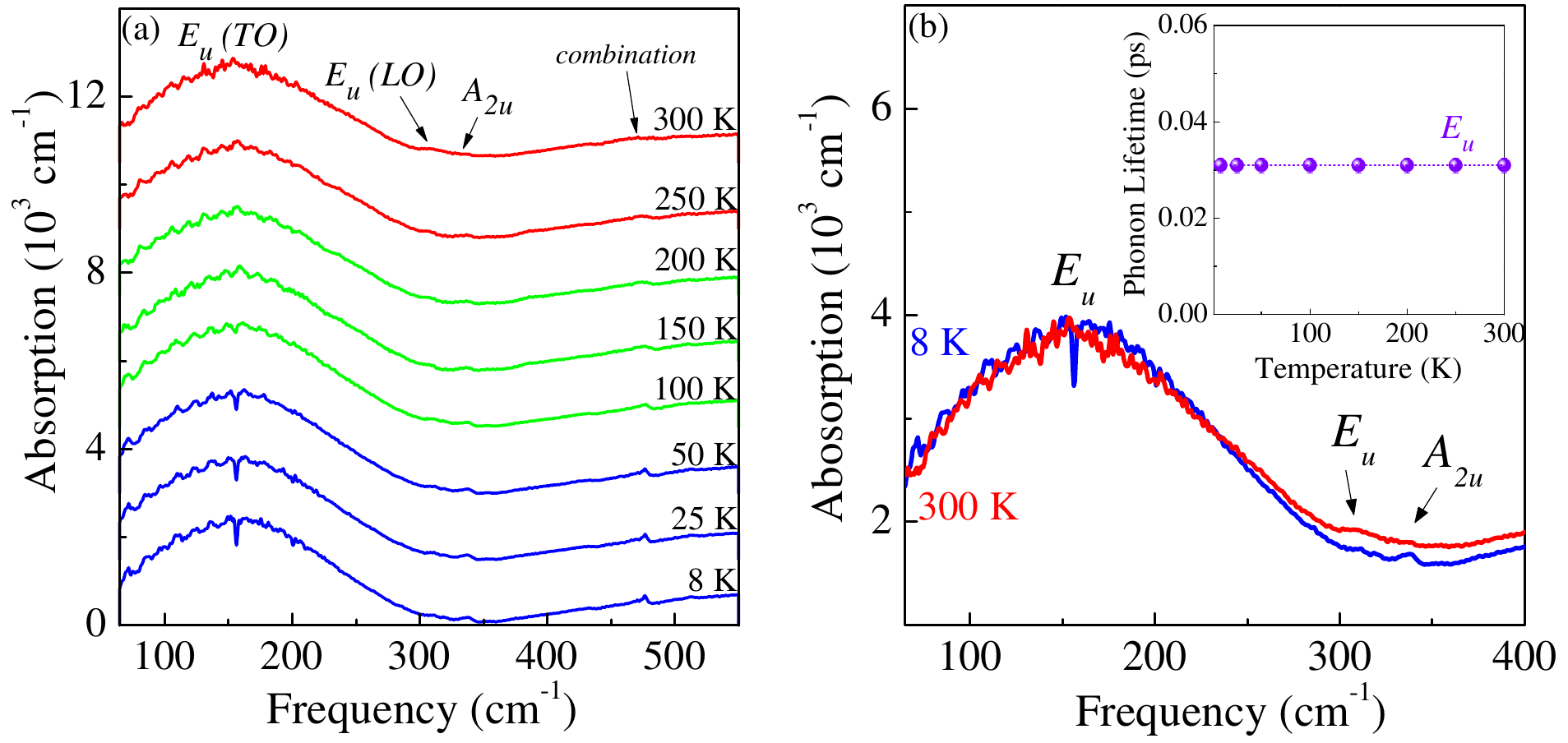}
\caption{{\bf{\label{Infrared} Infrared absorption studies of 1T-HfS$_2$.}} (a) Infrared absorption spectra of 1T-HfS$_2$ as a function of temperature. The  color scheme emphasizes the different phases, and the curves are offset for clarity. (b) Close-up view of the infrared response of 1T-HfS$_2$ on an absolute scale. The inset displays the phonon lifetime of the $E_{u}$ mode as a function of temperature.  Error bars are on the order of the size of the data points. 
}
\end{figure*}

Figure \ref{Infrared}(a) summarizes the infrared response of single crystalline 1T-HfS$_2$.
We assign the vibrational modes, symmetries, and displacement patterns based upon prior literature as well as our complementary lattice dynamics calculations [Table I] \cite{Ibanez2018, Najmaei2018, Roubi1988, Cingolani1987}. 
There are two fundamental infrared-active phonons. 
The $E_u$ symmetry mode is extremely broad and centered at 155 cm$^{-1}$. It is ascribed to an in-plane, out-of-phase motion of the sulfur layers against the hafnium. As we will discuss below, the maximum corresponds to the transverse optical ($TO$) phonon frequency. In the bulk form peaks in absorption always correspond to the $TO$ component of the mode. The $A_{2u}$ mode at 336 cm$^{-1}$ is much weaker and narrower. This feature is due to an out-of-plane + in-phase sulfur layer stretching, and although it is polarized in the $c$ direction, it appears in nascent form here probably due to surface flakiness or slight misalignment. In any case, it is very small. There is also a minute structure near 310 cm$^{-1}$  which, as we shall see below, corresponds to the weakly activated longitudinal optical ($LO$) frequency of the $E_u$ mode. Of course, each polar phonon has separate $LO$ and $TO$ components, and the $LO$ frequency is always higher than that of the $TO$ frequency due to the polarizability of the surrounding medium. Displacement patterns for each phonon are shown in Fig.~\ref{fig:dft}(c) and summarized in Table 1.

Figure \ref{Infrared}(a) also displays the infrared response of 1T-HfS$_2$ as a function of temperature. 
What is striking about these results is the overall lack of temperature-induced change over the 300 - 8 K range -  in both peak position and intensity. Traditionally, modeling of frequency vs. temperature effects provides important information on anharmonicity in a solid as well as the various force constants. 
The idea is to bring together frequency vs. temperature plots along with one of several different equations that depend upon the situation.  The absence of mode hardening with decreasing temperature suggests, however, that anharmonic effects in 1T-HfS$_2$ are modest and that energy scales are high. Oscillator strength sum rules are obeyed, as expected.
Below 100 K, there is a small resonance in the form of a dip that develops on top of the 155 cm$^{-1}$ $E_u$ symmetry phonon. This structure is one of the spectroscopic signatures of a weak local lattice distortion. Because evidence for the effect is stronger in the Raman response, we will defer this discussion until later. Partial sum rules on the $E_u$ vibrational mode are obeyed, as expected.

Another way to reveal temperature effects is to examine phonon lifetimes [Fig. \ref{Infrared}(b)]. These values, which are expressions of the Heisenberg uncertainty principle,  can be calculated from the linewidth of each vibrational mode as 
$  \tau = \frac{\hbar}{\Gamma}$,
where $\Gamma$ is the full width at half maximum and $\hbar$ is the reduced Planck's constant \cite{Wooten1972}. We find  that the phonon lifetime of the 
$E_u$ mode is 0.03 ps -  exceptionally short compared to that of the $A_{2u}$ mode. The lifetime of the $E_u$ mode is nearly insensitive to temperature as well, meaning that this mode dissipates energy very well - even at low temperature. In other words, $\tau$ is large because there are many scattering events. On the other hand, the lifetime of the 336 cm$^{-1}$ $A_{2u}$ feature is on the order of 0.4 ps. Overall, these phonon lifetimes are shorter than those of 2H-MoS$_2$ ($\tau$ = 5.5 and 2.3 ps for the $E_{1u}$ and A$_{2u}$ modes, respectively) as well as those of polar semiconductors like GaAs, ZnSe, and GaN (which tend to be between 2 and 10 ps)  \cite{Sun2013}. Likewise, silicon has a phonon lifetime between 1.6 and 2 ps depending on the carrier (hole) density \cite{Letcher2007}. This means that carrier-phonon scattering is of greater importance in 1T-HfS$_2$ than in 2H-MoS$_2$ or the traditional semiconductors.  Employing a characteristic phonon velocity of 4700 m/s \cite{Kang2015}, we find mean free paths in 1T-HfS$_2$ of 143 pm and 1740 pm for the $E_{u}$ and $A_{2u}$ modes, respectively. The mean free path for the $E_{u}$ mode is slightly smaller than all of the characteristic length scales in the system including the 369 pm van der Waals gap, the 289 pm sheet thickness, and the 253 pm Hf--S bond length whereas the mean free path for the $A_{2u}$ mode is 4 or 5 times larger than these characteristic length scales.

\subsection*{Revealing the Born effective charge via infrared spectroscopy}

The Born effective charge of transition metal dichalcogenides has been of sustained interest \cite{Iwasaki1982,  Uchida1978, Lucovsky1973, Sun2013, Sun2009}. This is because Born  charge can be calculated from first principles as summarized in the previous section and revealed directly from spectroscopic data by taking into account the relationship between the longitudinal and transverse optic phonon frequencies as indicated in Equation \ref{eq:LO-TO2}.

As a reminder,  $\omega_{LO}$ and $\omega_{TO}$ are the longitudinal and transverse optic phonon frequencies, $c$ is the speed of light, $N$ is the number of formula units in the unit cell, $V$ is the volume of the 1T-HfS$_2$ formula unit, $e$ is the electronic charge, $\epsilon_0$ is the permittivity of free space, $\varepsilon$($\infty$) is the dielectric constant after the phonons, and $m_k$ is the effective mass.  Employing Equation~\ref{Mass}, we find $m_k$ = 47.02 u for 1T-HfS$_2$.  We also use $\omega_{LO}$ = 310 cm$^{-1}$,  $\omega_{TO}$ = 155 cm$^{-1}$,  $\varepsilon$($\infty$) = 6.20, $N$ = 1, and $V$ = 66.44 \AA$^3$. In the absence of a robust experimentally determined volume, we used a theoretically predicted value\cite{Lucovsky1973}. $Z^*_{\rm{B}}$ is extremely sensitive to the value of $m_k$ and the choice of $\varepsilon$($\infty$). Employing these numbers, we find $Z^*_{\rm{B}}$ = 5.3$e$.

Interestingly, prior studies have led to a variety of Born effective charges for HfS$_2$ with values from 3.46$e$ to 5.5$e$ \cite{Uchida1978, Lucovsky1973, Chen2016}. These findings are summarized in Table 2. 
The variations in $Z^*_{\rm{B}}$  are  due to  differences in $m_k$, cell volume, and (to a lesser extent) ${\epsilon}{(\infty)}$ as well as variability  ${\omega}_{LO}$ and ${\omega}_{TO}$. Our  value of 5.3$e$  is consistent with the large  $LO$-$TO$ splitting which signals robust ionicity. By comparison, transition metal dichalcogenides like 2H-MoS$_2$ have much smaller LO-TO splitting and  $Z^*_{\rm{B}}$ =  1.11$e$ in the $ab$-plane \cite{Sun2009, Wieting1971}.

Born effective charge can be decomposed into polarizability ($\alpha$) and local effective charge ($Z^{*}$) as: 

\begin{equation}
    \alpha = \frac{(\varepsilon(\infty)-1)}{N[1-(1-\varepsilon(\infty))\frac{\eta}{V}]}, 
\end{equation}

and 

\begin{equation}
    Z^* = Z^{*}_B[1-\eta(\frac{N\alpha}{V})].
\end{equation}

\noindent
Here, $\eta$ is the depolarization constant. For two-dimensional systems, $\eta$ = 1/3 \cite{Carr1985, Kittel2004}. We find $\alpha$ = 5.07 $\AA^3$ and $Z^*$ = 5.2$e$ from our experimental value of the Born charge in the ${ab}$ plane. These values compare well with those obtained from our theoretical value of $Z_B^*$:  $\alpha$ = 6.85 $\AA^3$ and $Z^*$ = 6.2$e$. These results along with  literature values are summarized in Table \ref{Table 3}.

Polarizability entails the sum of all cationic and anionic electron cloud volume contributions whereas local effective charge is related to short range interactions indicative of ionic displacement and chemical bonding. These values reveal the nature of chemical bonding.   Typically, covalent systems such as silicon have $Z^*$ close to zero whereas  ionic materials have larger local effective charges. For example,  $Z^*$ = 0.15$e$ for MoS$_2$ \cite{Sun2009}, $Z^*$ = 1.14$e$ for MnO \cite{Sun2011}, and $Z^*$ = 0.8$e$ for NaCl \cite{Ashcroft1976}. Higher effective and ionic charge in MnO is consistent with the distinction of a Reststrahlen band. 1T-HfS$_2$ clearly has the largest local charge in this series.

\subsection*{Origin of anomalous Born effective charge}

In a completely ionic crystal where electrons are attached to ions and  displaced along with them by the same exact amount, the dynamical Born effective charge is equal to the formal ionic charge - which  would have given $Z^*_{\rm{B}}$ for Hf.  1T-HfS$_2$ is closer to this limit due to the low electronegativity of the cation (1.3 in the Pauling scale). This is lower than any 3$d$ transition metal, including Mo which has an electronegativity of 2.2. The electronegativity of S is 2.6, so the Mo-S bonds in MoS$_2$ are highly covalent. Nevertheless, we note that the Born effective charge of Hf in the in-plane direction is more than 50\% larger than the nominal charge of +4. This anomalous Born charge signals either (i) covalency between the cations and  anions or (ii) cation  polarizability \cite{ghosez1998}. Uchida explored the issue in terms of static and dynamic charge \cite{Uchida1978}. We can address the question more robustly with contemporary tools.

Maximally localized Wannier functions can be utilized to explain the origin of anomalous Born effective charges \cite{Ghosez2000}. The macroscopic electronic polarization can be expressed in terms of the center of localized Wannier functions as
\begin{equation}
    P_\beta^{el} = \frac{1}{\Omega_0} \sum_{n}^{occ} \int r_\beta |W_n(r)|^2 d^3r,
    \label{eq:wannier}
\end{equation}
where $W_n(r)$ is the Wannier function and the sum is over the filled Wannier orbitals. By displacing the Hf atoms in the in-plane and out-of-plane directions, it is possible to calculate the shift of the center of each Wannier function, and hence get a orbital-by-orbital or band-by-band decomposition of the Born effective charges.

Table~ \ref{tab:wannier} shows the band-by-band contributions to the Born effective charges of Hf. It turns out the despite the large electronegativity difference between Hf and S, covalency of Hf-S bonds is the dominant reason behind the anomalous Born effective charge in this compound: the Wannier centers of the electrons in the S-3$p$ bands displace significantly when the Hf ion is displaced. On the other hand, the Hf-5$p$ electrons are displaced almost exactly as much as the Hf ion core itself. Thus, S-3$p$ orbitals contribute most to the difference between Born effective and nominal charge.

Taking a closer look at the electrons in the S-3$p$ orbitals, Fig.~\ref{fig:wannier} shows one of the Wannier functions with and without the Hf ion displacement. In line with the strongly ionic density of states, the S-$p$ electrons are mostly localized on S, with small lobes on Hf indicating hybridization. When Hf atoms are displaced in an in-plane direction, one of the three Hf-S bonds is shortened while the other two are elongated. The shortened bond causes the Wannier function to be tilted, and its center shifted towards the Hf atom, shown in Fig.~\ref{fig:wannier}(b). This explains why S-3$p$ orbitals will contribute a positive polarization value when Hf atoms are moving in-plane. On the other hand, when Hf atoms move in the out-of-plane direction, the shape of the hybridized electron density on the Hf atom changes, showing a significant qualitative change in the hybridization, as well as a shift in the S-3$p$ Wannier centers parallel to the Hf displacement. This leads to a negative dynamic contribution of S-3$p$ to the out-of-plane Hf effective charge. Though only one S-3$p$ orbital is shown here, the others are similar.

As a comparison, the Born effective charge of Mo in 2H-MoS$_2$ is 1.1 - 1.2$e$ in the in-plane direction \cite{Sun2009}, and theoretical results suggest a sign reversal of the cation Born charge \cite{Nicholas2017}. This is likely because of a much stronger covalency in MoS$_2$ than in HfS$_2$, which results in more electrons transferred from S anions onto Mo cations when Mo ions are displaced. Note that there is a $4d_{z^2}-p_z$ antibonding orbital in Mo-S bonds near the Fermi level because Mo's 4$d$ band is partially filled, and this orbital can result in large electron transfer in a way similar to $\pi$-backbonding in organic chemistry \cite{Nicholas2017}. Our DFT calculation confirms the partially filled band structure as shown in the Supplementary Information.

\subsection*{Raman scattering spectroscopy reveals weak structural crossovers in 1T-HfS$_2$}

Figure \ref{Raman}(a) summarizes the Raman scattering response of  1T-HfS$_2$  as a function of temperature.
We assigned the spectral features based upon our lattice dynamics calculations and prior literature [Table \ref{Table 1}] \cite{Ibanez2018, Roubi1988,Cingolani1987}.  
The two Raman-active fundamentals at 259 and 336 cm$^{-1}$ are even symmetry vibrational modes. We attribute these features to $E_g$ symmetry (in-plane, out-of-phase sulfur layer shearing)  and the $A_{1g}$ symmetry (out-of-plane opposing sulfur motions leading to layer breathing), respectively. There is also a weak overtone mode near 650 cm$^{-1}$ that is slightly less than twice the frequency of the $A_{1g}$ fundamental. This feature has also been referred to as a second order mode \cite{Roubi1988}. There are several hybrid features as well. For instance,
the $E_u$ ($TO$) mode  appears weakly in the Raman spectrum  near 132 cm$^{-1}$ due to an in-plane motion of the sulfur against the hafnium.  
Finally, the  shoulder near 325 cm$^{-1}$ has been the subject of some controversy in the literature. It was previously described as an in-plane shearing ($E_u$), an out-of-plane sulfur translation (A$_{2u}$), or even a surface phonon \cite{Najmaei2018, Ibanez2018, Roubi1988, Cingolani1987}. 
Based upon prior pressure studies, the position of this phonon diverges from that of the A$_{1g}$ mode near 325 cm$^{-1}$ as pressure is increased \cite{Ibanez2018}. If this feature was a surface phonon it would likely track parallel to the much larger mode. Because it does not, it is unlikely to be a surface phonon. With the help of prior literature as well as the common position with the $E_u$ mode in the infrared studies, we assign this structure as an $E_u$ symmetry phonon \cite{Najmaei2018, Ibanez2018, Roubi1988, Cingolani1987}. Additional justification for this assignment comes from our phonon density of states calculations and the predicted order of the hybrid modes around the $A_{1g}$ fundamental [Fig.~\ref{fig:dft}(b)].  We do not see the hybrid $A_{2u}$ mode in our spectra - even at low temperature, although different laser wavelengths should reveal it \cite{Roubi1988}.

Figures \ref{Raman}(b) and (c) display  close-up views of the  $A_{1g}$ and $E_g$ symmetry Raman-active fundamentals as a function of temperature in contour form. 
While there is little frequency sensitivity in either feature (due to the high energy scales in this system), there are noticeable linewidth effects. 
Focusing on the behavior of the $A_{1g}$ layer expansion mode in  Fig. \ref{Raman}(b), we see that the linewidth narrows considerably with decreasing temperature, with slight broadening across $T_1$ $\approx$ 60 K and $T_2$ $\approx$ 210 K. There is also a slight redshift across $T_1$. 
 Analysis of the $E_g$ sulfur-layer stretching mode in Fig. \ref{Raman}(c), again shows linewidth narrowing as base temperature is approached, also with noticeable broadening across the two crossover regimes. 
A slight redshift below $T_1$ is again present. We attribute the 60 and 210 K transitions in 1T-HfS$_2$ to local lattice distortions involving a slight motion of the S centers with respect to the Hf ions so as to change the bond lengths and angles a little while maintaining the same overall space group.

Using these linewidth trends and the technique described previously, we calculated phonon lifetimes for the Raman-active vibrational modes of 1T-HfS$_2$ [Fig. \ref{Raman}(d)]. As a reminder, the $A_{1g}$ and $E_g$ modes are the fundamentals. The lifetime of the $A_{1g}$ phonon rises steadily with decreasing temperature. The behavior of the $E_g$ symmetry mode is different. It rises gradually below $T_2$ and dramatically across $T_1$.  This suggests that carrier-phonon scattering is reduced with decreasing temperature. Overall the lifetimes of the Raman-active even symmetry modes are similar to those of the infrared-active phonon modes in 1T-HfS$_2$ [Fig. \ref{Infrared}(b)] - with the exception of the  $E_u$  symmetry vibrational mode which is very lossy and therefore sports an extremely short lifetime. 
This is one surprising feature in 1T-HfS$_2$ that is not replicated in more traditional systems like 2H-MoS$_2$.

\subsection*{Toward chemical bonding in nanoscale analogs of 1T-HfS$_2$}

Overall, there has been surprisingly little discussion of chemical bonding and structural phase transitions in transition metal dichalcogenides in the literature. Much of the focus has instead been on oxides - particularly ferroelectrics - where questions of charge, polarizability, hybridization, and dynamic covalency are considered to be ``solved problems". The mechanism  of polar displacement-induced charge transfer is, however, under-explored and may be responsible for enhanced Born charge compared to the nominal charge in oxides as well. This is not the case in transition metal dichalcogenides or complex chalcogenides, and there is much discrepancy in the literature - with estimates of simple quantities like Born charge varying by as much as 40\% and almost no attention paid to the role of spin-orbit coupling in heavy chalcogenides. This work places the properties of 1T-HfS$_2$ on a firm foundation for understanding (and even actuating) functionality under external stimuli. It also illustrates a broader approach to exploring chemical bonding in the chalcogenides and opens the door to examination of nanoscale analogs such as tubes, particles, and sheets as well as effects of $n$- and $p$-type doping and intercalation.

\section*{Methods}

 1T-HfS$_2$ single crystals were grown via chemical vapor transport by 2D Semiconductors, Inc. Prior to traditional infrared and Raman scattering measurements, the sample was surface-exfoliated to remove surface impurities and then mounted on a round pin-hole aperture exposing the $ab$-plane.  Far infrared studies were performed using a Bruker IFS 113V Fourier-infrared spectrometer equipped with a bolometer detector covering the 20-700 cm$^{-1}$ frequency range with 2 cm$^{-1}$ resolution. The measured transmittance was converted to absorption as $\alpha$($\omega$)= -$\frac{1}{\textit{d}}$ln($\mathcal{T}$($\omega$)), where  $\textit{d}$ is the thickness (in this case, approximately 8.4 $\times$ 10$^{-4}$ cm). Raman scattering spectroscopy was carried out using a LabRAM HR Evolution Raman spectrometer 
(50-750 cm$^{-1}$) using an excitation wavelength of 532 nm at a power of 0.1 mW with an 1800 line/mm grating. An open flow cryostat was used to control temperature between 4 and 300 K. Traditional peak fitting techniques were employed as appropriate.

First principles calculations were performed using the projector augmented wave approach as implemented in the Vienna Ab-initio Simulation Package ({\sc vasp})\cite{Blochl1994, Kresse1996, Kresse1999}. Both the Born effective charges and the high frequency dielectric constant are determined from the response to finite electric fields. The high-frequency dielectric constant $\epsilon(\infty$), which only contains the electronic contribution, is obtained by differentiating the polarization with respect to the external electric field with clamped ions:

\begin{equation}
    \epsilon_{ij}^\infty = \delta_{ij} +\frac{4\pi}{\epsilon_0}\frac{\partial P_i}{\partial \mathcal{E}_j},
\end{equation}
where the polarization is calculated using the implementation of the the Perturbation Expression After Discretization (PEAD) approach in the VASP package \cite{PEAD1,PEAD2}. Electric fields of 2 meV$/$\AA, 2 meV$/$\AA, and 10 meV$/$\AA ~are applied separately along $a$, $b$, and $c$ axes to calculate the derivatives. Similarly, the Born effective charge is calculated through the derivatives of the polarization with respect to the ionic displacements. At this step, using the Hellman-Feynman forces enables the use of the more computationally efficient formula 
\begin{equation}
    Z_{ij}^*=\frac{\Omega}{e}\frac{\partial P_i}{\partial u_j} 
    = \frac{1}{e}\frac{\partial^2 \mathcal{F}}{\partial u_j \partial \mathcal{E}_i} = \frac{1}{e}\frac{\partial F_j}{\partial \mathcal{E}_i}.
\end{equation}
Here $\mathcal{F}$ is the electric enthalpy which is the sum of the Kohn-Sham energy and the energy gain due to the interaction between the polarization and the external electric field: $\mathcal{F} = E_{KS} - \Omega\mathbf{P}\cdot\mathbf{\mathcal{E}}$. The Hellman-Feynman force $F$ is given by $F_i=\frac{\partial \mathcal{F}}{\partial u_i}$. $F$  solely depends on the ground state wavefunction, and hence is easier to calculate than the polarization. As a reliability check, density functional perturbation theory (DFPT) \cite{Baroni1987} combined with multiple functionals (including PBEsol\cite{PBEsol} and revised-TPSS meta-GGA\cite{metaGGA}) with spin-coupling is also performed to get the Born effective charge, which can be found in the Supplementary Information. Both approaches provide similar results for the dielectric constant and Born effective charge. We chose the latter method for its compatibility with Hatree-Fock method, and as a result, the hybrid functionals.

All the first-principles calculations are performed in the primitive 3 atom unit cell with a $12 \times 12 \times 6 $ k-point grid and cut-off energy of 500 eV. The lattice constants and vectors are taken from the experimental literature, but the internal coordinates of the S ions are obtained through structural optimization of forces. The energy tolerance for self-consistency is set to $10^{-8}$ to get a well-converged wavefunction. To reproduce the experimental bandgap more closely, HSE hybrid functional is employed \cite{heyd2003hybrid}. In the case of 1T-HfS$_2$, a energy band gap of 2.05 eV can be achieved by using the screening parameter of 0.2, which is the so-called HSE06 approximation. Reports of the band gap of HfS$_2$ span values from 1.96 eV (from optical absorption) \cite{greenaway1965preparation} to 2.85 eV (from combined angle-resolved and inverse photoemission) \cite{traving2001combined}. Note that band structures calculated using PBEsol or meta-GGA functionals both underestimate the band gap by at least a factor of two, which  influences the prediction accuracy of electric field response. A comparison between different functionals is presented in the Supplementary Information. Since Hf is a heavy element with strong spin-orbital coupling (SOC) \cite{yagmurcukardes2019raman}, DFT calculations that take SOC into account were also performed, but no significant change of Born effective charges and phonon frequencies are observed. A detailed comparison of all theoretical results is provided in the Supplementary Information.

To further explain the origin of Born effective charge, we employed the maximum localized Wannier function (MLWF) \cite{Marzari1997, Marzari2012}  to project the  band-by-band contribution. The Wannier90 software package is used for this analysis \cite{MOSTOFI20142309}.

\label{sec:method}

\section*{Acknowledgements}

Research at the University of Tennessee is supported by the U.S.
Department of Energy, Office of Basic Energy Sciences, Materials
Science Division under award DE-FG02-01ER45885. The work 
at the University of Minnesota is supported primarily by the National Science Foundation through the University of Minnesota MRSEC under Award Number DMR-2011401. We acknowledge the Minnesota Supercomputing Institute (MSI) at the University of Minnesota for providing resources that contributed to the research results reported within this paper. We thank S. Najmaei and I. Boulares at the U. S. Army Research Lab for the 1T-HfS$_2$ crystal and useful conversations.

\section*{Contributions}

This project was conceived by JLM and SNN. Raman scattering and infrared absorption measurements were conducted by SNN and data were analyzed by SNN and JLM. Theoretical studies and density functional calculations were carried out by SL and TB. all authors discussed data. The manuscript was written by SNN, SL, TB, and JLM. All authors read and commented on the manuscript.

\section*{Competing interests}
The authors declare no competing interests.

\clearpage
\section{Figures}

\begin{figure*}
\centering
\includegraphics[width = 6.0in]{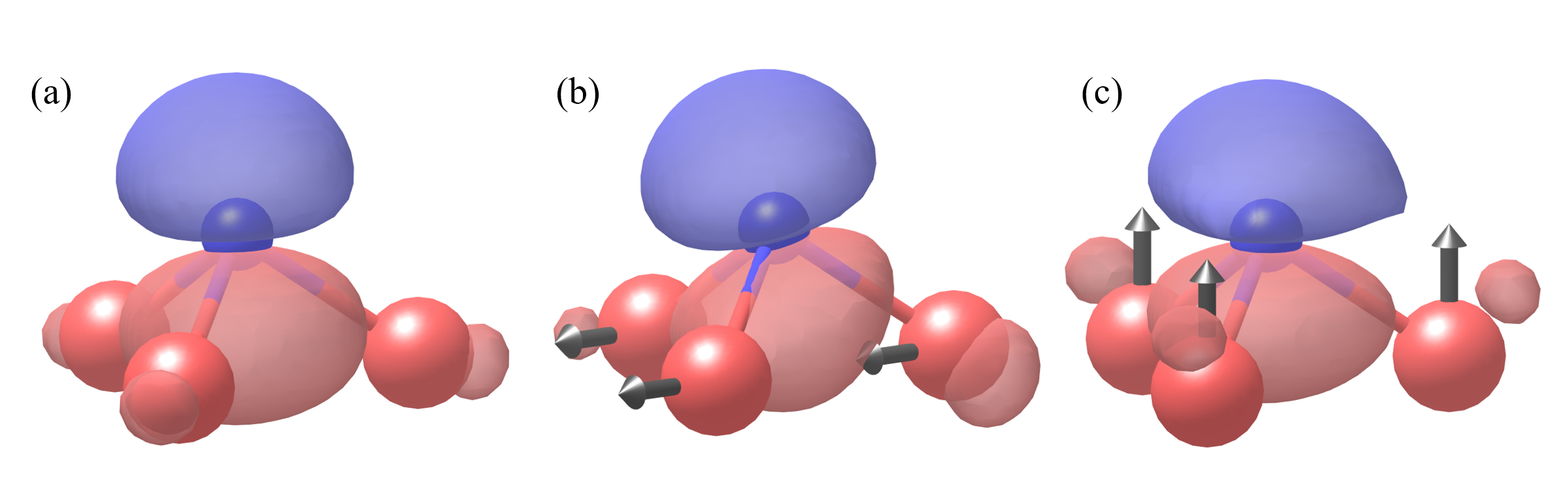}
\caption{{\bf{Visualization of sulfur $p_z$ maximally localized Wannier orbital.}}\label{fig:wannier} Visualization of a sulfur $p_z$ maximally localized Wannier orbital. The blue atom is a sulfur center whereas red atoms are bonded hafnium centers. The red and blue lobes of the orbital indicates the opposite signs of the wavefunction. No structural distortion is present in (a). Hf atoms are displaced in-plane and out-of-plane in (b) and (c) respectively.}

\end{figure*}

\begin{figure*}
\centering
\includegraphics[width = 6.0in]{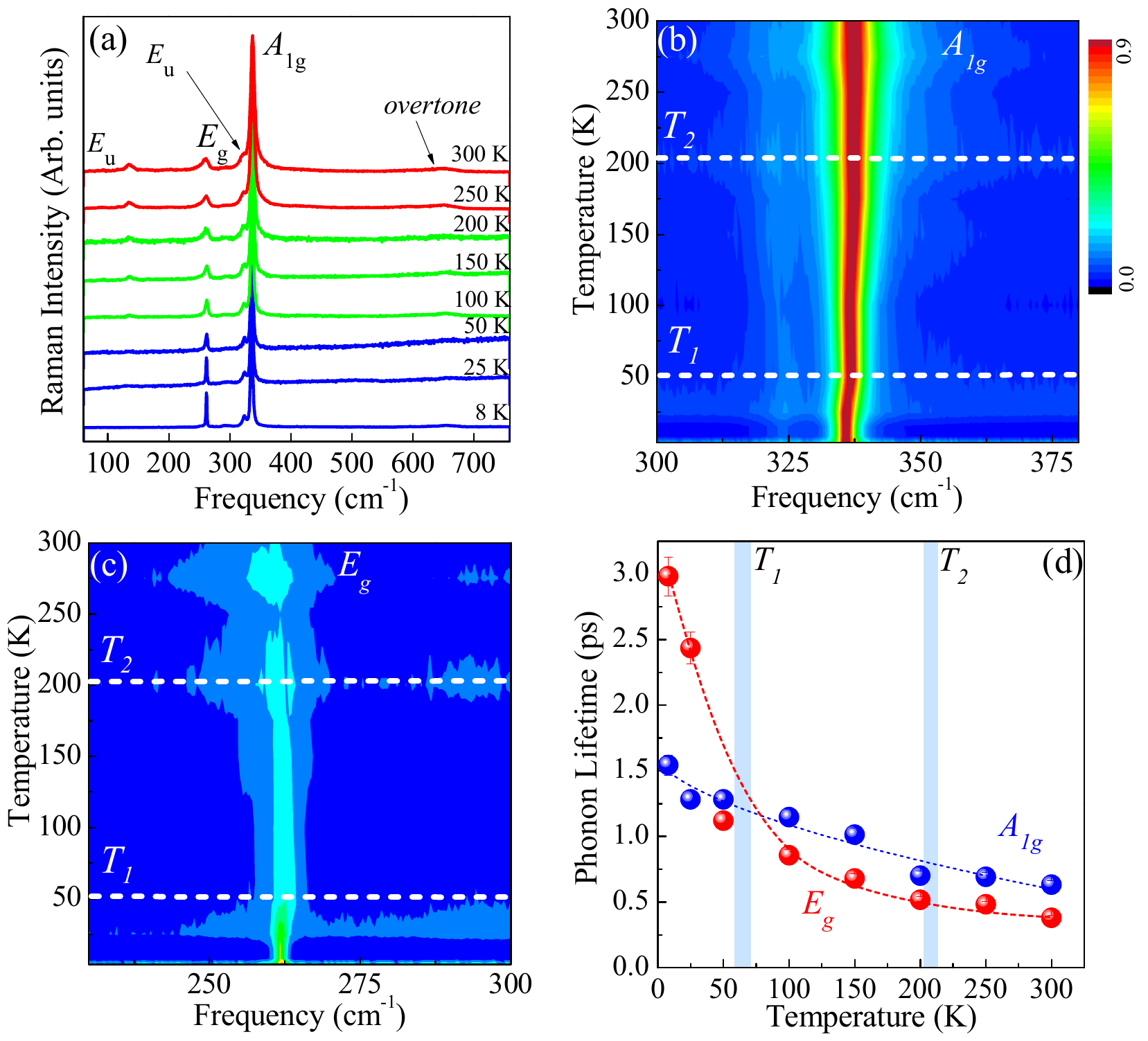}
\caption{{\bf{\label{Raman} Variable temperature Raman spectroscopy and associated phonon lifetime analysis of 1T-HfS$_2$.}} (a) Raman scattering response of 1T-HfS$_2$ as a function of temperature. The mode symmetries are labeled, and the spectra are offset for clarity. (b) Contour plot showing  subtle changes in the $A_{1g}$ mode as a function of temperature. Weak structural distortions at $T_1$ and $T_2$ are indicated. The color scale give relative intensity. (c) Contour plot showing the development of the $E_g$ mode with temperature. (d) Phonon lifetime of two Raman-active modes as a function of temperature. Data is available in the Supplementary Information $T_1$ and $T_2$ are denoted by  the color scheme in (a) and the vertical blue lines in (d).
}
\end{figure*}

\clearpage
\section{Tables}

\begin{table}[bth]
\centering
\caption{Vibrational mode assignments of 1T-HfS$_2$. All values are in cm$^{-1}$} 
\begin{tabular}{ccccc}
\hline\hline
\textbf{Theoretical} & \textbf{Symmetry} & \textbf{Infrared} & \textbf{Raman} & \textbf{Mode Displacement}\\
\hline

152 & $E_u$ ($TO$) & 155 & 132 &  in-plane, out-of-phase motion \\ 
& & & & of sulfur against hafnium\\
\hline
259 & $E_g$ & - & 259  & in-plane, out-of-phase \\ 
& & & & sulfur layer shearing\\
\hline
300 & $E_u$ ($LO$) & 310 & 325 &  in-plane, out-of-phase motion \\ 
& & & & of sulfur against hafnium\\
\hline
316 & $A_{1g}$ & - & 336 &  out-of-plane, out-of-phase \\ 
& & & & sulfur layer breathing\\
\hline
321 & A$_{2u}$ & 336 & - &  out-of-plane, in-phase \\ 
& & & & stretch of sulfur layer\\
\hline
\hline 
   & $E_u$ + $A_{2u}$ & 475 &   & combination mode\\
\hline
 & 2($A_{1g}$) &  & 650   & overtone mode\\
\hline
\hline
\end{tabular}
\vspace{0.05in}

\label{Table 1}
\end{table}

\begin{table*}[tbh]
\centering
\caption{Comparison of parameters and Born effective charge for the $E_u$ mode of 1T-HfS$_2$. Literature results and our own work - both experimental and theoretical - are included.} 

\begin{tabular}{c@{\hskip \tabcolsep}c@{\hskip \tabcolsep}c@{\hskip \tabcolsep} c@{\hskip 2\tabcolsep} c@{\hskip 2\tabcolsep} c@{\hskip 2\tabcolsep} c@{\hskip 2\tabcolsep} }
\hline\hline
  $\omega$({\bf{$LO$}}) (cm$^{-1}$) & $\omega$(\textbf{$TO$}) (cm$^{-1}$) &  {\textbf{$\varepsilon$($\infty$)}} & $m_k$ (u) & $V$ ($\AA^3$) &{Z$^{*}_B$} ($e$)& Reference \\
\hline
%
%
300 & 152 & 6.33 & - & - & 3.46  & Chen \textit{et. al.} \cite{Chen2016} \\
318 & 166 & 6.20 & 23.59 & 69.46 & 3.90 & Lucovsky \textit{et. al.} \cite{Lucovsky1973}\\
321 & 166 & 6.20 & -& - & 5.50 & Uchida \textit{et. al.} \cite{Uchida1978}\\
\hline
310 & 155 & 6.20 & 47.02 & 66.44 & 5.3 & This work (experiment)\\
300  & 152  &  8.09 & 47.02  &  66.44 & 6.4 & This work (theory)\\
\hline
\hline
\end{tabular} 
\vspace{0.05in}
\label{Table 2}
\end{table*}

\begin{table}[tbh]
\centering
\caption{Comparison of Born charge, polarizability, and local (ionic) charge from the $E_u$ mode of 1T-HfS$_2$. Literature values and our own work are included.} %
\begin{tabular}{c@{\hskip 2\tabcolsep}c@{\hskip 2\tabcolsep} c@{\hskip 2\tabcolsep} c@{\hskip 2\tabcolsep}   }
\hline\hline
   Z$^*_B$ ($e$) & $\alpha$ ($\AA^3$) & Z$^*$ ($e$) & Reference \\
\hline
- & 2.2  & 0.86 & Iwasaki \textit{et. al.} \cite{Iwasaki1982}\\
3.46 & - & - & Chen \textit{et. al.} \cite{Chen2016} \\
3.90 & - & - & Lucovsky \textit{et. al.} \cite{Lucovsky1973}\\
5.50 & - & 2.60 & Uchida \textit{et. al.} \cite{Uchida1978}\\
\hline
5.3 & 5.07 & 5.2 & This work (experiment)\\
6.4 & 6.85 & 6.2 & This work (theory)\\
\hline
\hline
\end{tabular} 
\vspace{0.05in}
\label{Table 3}
\end{table}

\begin{table}
   \caption{A band-by-band decomposition of Born effective charge of Hf in HfS$_2$ using integration of Wannier function. All units are in $e$. The core charges correspond to the charge of the ionic Hf core of the PAW potential used in the DFT calculation, which has all valence orbitals empty. (Only the 5$p$, 5$d$ and 6$s$ electrons of Hf atoms and 3$s$, 3$p$ electrons of S atoms are considered explicitly in the DFT calculation.) Hf-5$d$ orbitals are not shown because they are formally not occupied. Further details are available in the Supplementary Information.}
    \centering
        \begin{tabular}{c c c c c c}
        \hline \hline
          & \vtop{\hbox{\strut Core}\hbox{\strut charge}} & Hf:5$p$ & S:3$s$ & S:3$p$ & total\\
          \hline
         $Z_{B,xx}$& +10 & -6.08 & +0.30 & +2.34 & 6.56\\
         $Z_{B,zz}$& +10 & -6.20 & 0.01 & -1.86 & 1.93\\
         \hline \hline
    \end{tabular}
    \label{tab:wannier}
\end{table}



\clearpage


\begin{thebibliography}{10}

\bibitem{Kim2008}
Kim, B.~J., {\textit{et al}}.
\newblock {Novel J$_{eff}$=1/2 mott state induced by relativistic spin-orbit
  coupling in Sr$_2$IrO$_4$}.
\newblock \emph{Phys. Rev. Lett.} \textbf{101}, 076402 (2008).

\bibitem{Birol2015}
Birol, T. \& Haule, K.
\newblock {J$_{eff}$=1 /2 Mott-insulating state in Rh and Ir fluorides}.
\newblock \emph{Phys. Rev. Lett.} \textbf{114}, 096403 (2015).

\bibitem{Zwartsenberg2020}
Zwartsenberg, B., {\textit{et al}}.
\newblock {Spin-orbit-controlled metal–insulator transition in
  Sr$_2$IrO$_4$}.
\newblock \emph{Nat. Phys.} \textbf{16}, 290--294 (2020).

\bibitem{Cao2018}
Cao, G., {\textit{et al}}.
\newblock {Electrical control of structural and physical properties via strong
  spin-orbit interactions in Sr$_2$IrO$_4$}.
\newblock \emph{Phys. Rev. Lett.} \textbf{120}, 017201 (2018).

\bibitem{Cao2000}
Cao, G., {\textit{et al}}.
\newblock {Charge density wave formation accompanying ferromagnetic ordering in
  quasi-one-dimensional BaIrO$_3$}.
\newblock \emph{J. Phys. C} \textbf{113}, 657--662 (2000).

\bibitem{Chen2020}
Chen, Q., {\textit{et al}}.
\newblock {Realization of the orbital-selective Mott state at the molecular
  level in Ba$_3$LaRu $_2$O$_9$}.
\newblock \emph{Phys. Rev. Mater.} \textbf{4}, 064409 (2020).

\bibitem{Singleton2016}
Singleton, J., {\textit{et al}}.
\newblock {Magnetic properties of Sr$_3$NiIrO$_6$ and Sr$_3$CoIrO$_6$: Magnetic
  hysteresis with coercive fields of up to 55 T}.
\newblock \emph{Phys. Rev. B} \textbf{94}, 224408 (2016).

\bibitem{ONeal2019}
{O' Neal}, K.~R., {\textit{et al}}.
\newblock {Spin – lattice and electron – phonon coupling in 3d / 5d hybrid
  Sr$_3$NiIrO$_6$}.
\newblock \emph{npj Quant. Mater.} \textbf{48}, 1--6 (2019).

\bibitem{Ma2018}
Ma, Y., Kou, L., Huang, B., Dai, Y., \& Heine, T.
\newblock {Two-dimensional ferroelastic topological insulators in single-layer
  Janus transition metal dichalcogenides MSSe (M=Mo, W)}.
\newblock \emph{Phys. Rev. B} \textbf{98}, 085420 (2018).

\bibitem{Yao2018}
Yao, Q., Zhang, L., Bampoulis, P., \& Zandvliet, H.~J.
\newblock {Nanoscale investigation of defects and oxidation of HfSe$_2$}.
\newblock \emph{J. Phys. Chem. C} \textbf{122}, 25498--25505 (2018).

\bibitem{Zhu2017}
Zhu, H., {\textit{et al}}.
\newblock {Defects and surface structural stability of MoTe$_2$ under vacuum
  annealing}.
\newblock \emph{ACS Nano} \textbf{11}, 11005--11014 (2017).

\bibitem{ChenChen2017}
Chen, C., {\textit{et al}}.
\newblock {Surface phases of the transition-metal dichalcogenide IrTe$_2$}.
\newblock \emph{Phys. Rev. B} \textbf{95}, 094118 (2017).

\bibitem{Ideta2018}
Ideta, S., {\textit{et al}}.
\newblock {Ultrafast dissolution and creation of bonds in IrTe$_2$ induced by
  photodoping}.
\newblock \emph{Sci. Adv.} \textbf{4}, aar3867 (2019).

\bibitem{Pascut2014}
Pascut, G.~L., {\textit{et al}}.
\newblock {Dimerization-induced cross-layer quasi-two-dimensionality in
  metallic IrTe$_2$}.
\newblock \emph{Phys. Rev. Lett.} \textbf{112}, 086402 (2014).

\bibitem{Jiang2020}
Jiang, K., {\textit{et al}}.
\newblock {New pressure stabilization structure in two-dimensional PtSe$_2$}.
\newblock \emph{J. Phys. Chem. Lett.} \textbf{11}, 73427349 (2020).

\bibitem{Heine2020}
Kempt, R., Kuc, A., \& Heine, T.
\newblock {Two-dimensional noble-metal chalcogenides and phosphochalcogenides}.
\newblock \emph{Angew. Chem. Int. Ed.} \textbf{59}, 9242--9254 (2020).

\bibitem{Koz2020}
Kozhakhmetov, A., {\textit{et al}}.
\newblock {Scalable BEOL compatible 2D tungsten diselenide}.
\newblock \emph{2D Mater.} \textbf{7}, 015029 (2020).

\bibitem{Cheema2020}
Cheema, S.S., {\textit{et al}}.
\newblock {Enhanced ferroelectricity in ultrathin films grown directly on
  silicon}.
\newblock \emph{Nature} \textbf{580}, 478--482 (2020).

\bibitem{HfO2020}
Lee, H-J., {\textit{et al}}.
\newblock {Scale-free ferroelectricity induced by flat phonon bands in
  HfO$_2$}.
\newblock \emph{Science} \textbf{67}, 1--10 (2020).

\bibitem{Iwasaki1982}
Iwasaki, T., Kuroda, N., \& Nishina, Y.
\newblock {Anisotropy of lattice dynamical properties of ZrS$_2$ and HfS$_2$}.
\newblock \emph{J. Phys. Soc. Japan} \textbf{51}, 2233--22440 (1982).

\bibitem{traving2001combined}
Traving, M., {\textit{et al}}.
\newblock {Combined photoemission and inverse photoemission study of HfS$_2$}.
\newblock \emph{Phys. Rev. B} \textbf{63}, 035107 (2001).

\bibitem{Kanazawa2016}
Kanazawa, T., {\textit{et al}}.
\newblock {Few-layer HfS$_2$ transistors}.
\newblock \emph{Sci. Rep.} \textbf{6}, 22277 (2016).

\bibitem{Xu2015}
Xu, K., {\textit{et al}}.
\newblock {Ultrasensitive phototransistors based on few-layered HfS$_2$}.
\newblock \emph{Adv. Mater.} \textbf{27}, 7881--7887 (2015).

\bibitem{Najmaei2018}
Najmaei, S., {\textit{et al}}.
\newblock {Cross-plane carrier transport in van der Waals layered materials}.
\newblock \emph{Small} \textbf{14}, 1--11 (2018).

\bibitem{Roubi1988}
Roubi, L. \& Garlone, G.
\newblock {Resonance Raman spectrum of HfS$_2$ and ZrS$_2$}.
\newblock \emph{Phys. Rev. B} \textbf{37}, 6808--6812 (1988).

\bibitem{Uchida1978}
Uchida, S.-i. \& Tanaka, S.
\newblock {Optical phonon modes and localized effective charges of
  transition-metal dichalcogenides}.
\newblock \emph{J. Phys. Soc. Japan} \textbf{45}, 153--161 (1978).

\bibitem{Lucovsky1973}
Lucovsky, G., White, R.~M., Benda, J.~A., \& Revelli, J.~F.
\newblock {Infrared-reflectance spectra of layered group-IV and group-VI
  transition-metal dichalcogenides}.
\newblock \emph{Phys. Rev. B} \textbf{7}, 3859--3870 (1973).

\bibitem{Chen2016}
Chen, J.
\newblock {Phonons in bulk and monolayer HfS$_2$ and possibility of
  phonon-mediated superconductivity: A first-principles study}.
\newblock \emph{Solid State Commun.} \textbf{237-238}, 14--18 (2016).

\bibitem{Ibanez2018}
Ib{\'{a}}{\~{n}}ez, J., {\textit{et al}}.
\newblock {High-pressure Raman scattering in bulk HfS$_2$: comparison of
  density functional theory methods in layered MS$_2$ compounds (M = Hf, Mo)
  under compression}.
\newblock \emph{Sci. Rep.} \textbf{8}, 1--10 (2018).

\bibitem{Fu2017}
Fu, L., {\textit{et al}}.
\newblock {van der Waals epitaxial growth of atomic layered HfS$_2$ crystals
  for ultrasensitive near-infrared phototransistors}.
\newblock \emph{Adv. Mater.} \textbf{29}, 1700439 (2017).

\bibitem{Zhang2019}
Zhang, W., Netsu, S., Kanazawa, T., Amemiya, T., \& Miyamoto, Y.
\newblock {Effect of increasing gate capacitance on the performance of a
  p-MoS$_2$/HfS$_2$ van der Waals heterostructure tunneling field-effect
  transistor}.
\newblock \emph{Japn. J. Appl. Phys.} \textbf{58}, SBBH02 (2019).

\bibitem{Kang2015}
Kang, J., Sahin, H., \& Peeters, F.~M.
\newblock {Mechanical properties of monolayer sulphides: a comparative study
  between MoS$_2$, HfS$_2$, and TiS$_3$}.
\newblock \emph{Phys. Chem. Chem. Phys.} \textbf{17}, 27742--27749 (2015).

\bibitem{Singh2016}
Singh, D., Gupta, S.~K., Sonvane, Y., Kumar, A., \& Ahuja, R.
\newblock {2D-HfS$_2$ as an efficient photocatalyst for water splitting}.
\newblock \emph{Catal. Sci. and Technol.} \textbf{6}, 6605--6614 (2016).

\bibitem{photodetectorHf2018}
Wang, D., {\textit{et al}}.
\newblock {Selective direct growth of atomic layered HfS$_2$ on hexagonal boron
  nitride for high performance photodetectors}.
\newblock \emph{Chem. Mater.} \textbf{30}, 3819--3826 (2018).

\bibitem{Wu2017}
Wu, N., {\textit{et al}}.
\newblock {Strain effect on the electronic properties of 1T-HfS$_2$ monolayer}.
\newblock \emph{Physica E} \textbf{93}, 1--5 (2017).

\bibitem{Berwanger2020}
Berwanger, M., Ahuja, R., \& Piquini, P.~C.
\newblock {HfS$_2$ and TiS$_2$ monolayers with adsorbed C, N, P atoms: A first
  principles study}.
\newblock \emph{Catalysts} \textbf{10}, catal10010094 (2020).

\bibitem{Zheng2016}
Zheng, B., {\textit{et al}}.
\newblock {Vertically oriented few-layered HfS$_2$ nanosheets: Growth mechanism
  and optical properties}.
\newblock \emph{2D Mater.} \textbf{3}, 035024 (2016).

\bibitem{Lei2019}
Lei, C., {\textit{et al}}.
\newblock {Broken-gap type-III band alignment in WTe$_2$/HfS$_2$ van der Waals
  heterostructure}.
\newblock \emph{J. Phys. Chem. C} \textbf{123}, 23089--23095 (2019).

\bibitem{Kresse1999}
Kresse, G. \& Joubert, D.
\newblock From ultrasoft pseudopotentials to the projector augmented-wave
  method.
\newblock \emph{Phys. Rev. B} \textbf{59}, 1758--1775 (1999).

\bibitem{ghosez1998}
Ghosez, P., Michenaud, J.-P., \& Gonze, X.
\newblock {The physics of dynamical atomic charges: the case of ABO$_3$
  compounds}.
\newblock \emph{Phys. Rev. B} \textbf{58}, 6224--6240 (1998).

\bibitem{baroni2001phonons}
Baroni, S., De~Gironcoli, S., Dal~Corso, A., \& Giannozzi, P.
\newblock Phonons and related crystal properties from density-functional
  perturbation theory.
\newblock \emph{Rev. Mod. Phys.} \textbf{73}, 515--562 (2001).

\bibitem{Cingolani1987}
Cingolani, A., Lugar{\'{a}}, M., Scamarcio, G., \& L{\'{e}}vy, F.
\newblock {The Raman scattering in hafnium disulfide}.
\newblock \emph{Solid State Commun.} \textbf{62}, 121--123 (1987).

\bibitem{Wooten1972}
Wooten, F.
\newblock \emph{{Optical properties of solids}} (Academic Press, 1972).

\bibitem{Sun2013}
Sun, Q.~C., {\textit{et al}}.
\newblock {Spectroscopic determination of phonon lifetimes in rhenium-doped
  MoS$_2$ nanoparticles}.
\newblock \emph{Nano Lett.} \textbf{13}, 2803--2808 (2013).

\bibitem{Letcher2007}
Letcher, J., Kang, K., Cahill, D.~G., \& Diott, D.~D.
\newblock {Effects of high carrier densities on phonon and carrier lifetimes in
  Si by time-resolved anti-Stokes Raman scattering}.
\newblock \emph{Appl. Phys. Lett.} \textbf{90}, 252104 (2007).

\bibitem{Sun2009}
Sun, Q.~C., Xu, X., Vergara, L.~I., Rosentsveig, R., \& Musfeldt, J.~L.
\newblock {Dynamical charge and structural starin in inorganic fullerenelike
  MoS$_2$ nanoparticles}.
\newblock \emph{Phys. Rev. B} \textbf{79}, 205405 (2009).

\bibitem{Wieting1971}
Wieting, T.~J. \& Verble, J.~L.
\newblock {Infrared and Raman studies of long-wavelength optical phonons in
  hexagonal MoS$_2$}.
\newblock \emph{Phys. Rev. B} \textbf{3}, 4286--4292 (1971).

\bibitem{Carr1985}
Carr, G.~L., Perkowitz, S., \& Tanner, D.~B.
\newblock \emph{{Infrared and millimeter waves}}, vol.~13 (Academic Press,
  1985).

\bibitem{Kittel2004}
Kittel, C.
\newblock \emph{{Introduction to Solid State Physics}}.
\newblock 8$^{th}$ ed. (Wiley, 2004).

\bibitem{Sun2011}
Sun, Q.~C., Xu, X., Baker, S.~N., Christianson, A.~D., \& Musfeldt, J.~L.
\newblock {Experimental determination of ionicity in MnO nanoparticles}.
\newblock \emph{Chem. Mater.} \textbf{23}, 2956--2960 (2011).

\bibitem{Ashcroft1976}
Ashcroft, N.~W. \& Mermin, D.
\newblock \emph{{Solid State Physics}} (Thomson Learning, 1976).

\bibitem{Ghosez2000}
Ghosez, P. \& Gonze, X.
\newblock Band-by-band decompositions of the born effective charges.
\newblock \emph{J. Condens. Matter Phys.} \textbf{12}, 9179--9188 (2000).

\bibitem{Nicholas2017}
Pike, N.~A., {\textit{et al}}.
\newblock Origin of the counterintuitive dynamic charge in the transition metal
  dichalcogenides.
\newblock \emph{Phys. Rev. B} \textbf{95}, 201106 (2017).

\bibitem{Blochl1994}
Bl\"{o}chl, P.~E.
\newblock {Projector augmented-wave method}.
\newblock \emph{Phys. Rev. B} \textbf{50}, 17953 (1994).

\bibitem{Kresse1996}
Kresse, G. \& Furthm\"{u}ller, J.
\newblock {Efficient iterative schemes for ab initio total-energy calculations
  using plane-wave basis set}.
\newblock \emph{Phys. Rev. B} \textbf{54}, 11169--11186 (1996).

\bibitem{PEAD1}
Nunes, R.~W. \& Gonze, X.
\newblock Berry-phase treatment of the homogeneous electric field perturbation
  in insulators.
\newblock \emph{Phys. Rev. B} \textbf{63}, 155107 (2001).

\bibitem{PEAD2}
Souza, I., \'I\~niguez, J., \& Vanderbilt, D.
\newblock First-principles approach to insulators in finite electric fields.
\newblock \emph{Phys. Rev. Lett.} \textbf{89}, 117602 (2002).

\bibitem{Baroni1987}
Baroni, S., Giannozzi, P., \& Testa, A.
\newblock {Green's-function approach to linear response in solids}.
\newblock \emph{Phys. Rev. Lett.} \textbf{58}, 1861--1864 (1987).

\bibitem{PBEsol}
Csonka, G.~I., {\textit{et al}}.
\newblock Assessing the performance of recent density functionals for bulk
  solids.
\newblock \emph{Phys. Rev. B} \textbf{79}, 155107 (2009).

\bibitem{metaGGA}
Sun, J., {\textit{et al}}.
\newblock Self-consistent meta-generalized gradient approximation within the
  projector-augmented-wave method.
\newblock \emph{Phys. Rev. B} \textbf{84}, 035117 (2011).

\bibitem{heyd2003hybrid}
Heyd, J., Scuseria, G.~E., \& Ernzerhof, M.
\newblock Hybrid functionals based on a screened coulomb potential.
\newblock \emph{J. Chem. Phys.} \textbf{118}, 8207--8215 (2003).

\bibitem{greenaway1965preparation}
Greenaway, D.~L. \& Nitsche, R.
\newblock {Preparation and optical properties of group IV--VI$_2$ chalcogenides
  having the CdI$_2$ structure}.
\newblock \emph{J. Phys. Chem. Solids} \textbf{26}, 1445--1458 (1965).

\bibitem{yagmurcukardes2019raman}
Yagmurcukardes, M., Ozen, S., Iyikanat, F., Peeters, F.~M., \& Sahin, H.
\newblock {Raman fingerprint of stacking order in HfS$_2$- Ca(OH)$_2$
  heterobilayer}.
\newblock \emph{Phys. Rev. B} \textbf{99}, 205405 (2019).

\bibitem{Marzari1997}
Marzari, N. \& Vanderbilt, D.
\newblock {Maximally localized generalized Wannier functions for composite
  energy bands}.
\newblock \emph{Phys. Rev. B} \textbf{56}, 12847--12865 (1997).

\bibitem{Marzari2012}
Marzari, N., Fe, P., Yates, J.~R., \& Vanderbilt, D.
\newblock {Maximally localized Wannier functions : Theory and applications}.
\newblock \emph{Rev. Mod. Phys.} \textbf{84}, 1419--1475 (2012).

\bibitem{MOSTOFI20142309}
Mostofi, A.~A., {\textit{et al}}.
\newblock An updated version of wannier90: A tool for obtaining
  maximally-localised wannier functions.
\newblock \emph{Comput. Phys. Commun.} \textbf{185}, 2309 -- 2310 (2014).

\end{thebibliography}

\begin{thebibliography}{}

\bibitem{Wooten1972}
Wooten, F. {\it{Optical properties of solids}} (Academic Press, 1972).

\bibitem{baroni2001phonons}
Baroni, S., De Gironcoli, S., Dal Corso, A., and Giannozzi, P. Phonons and related crystal properties from density-functional perturbation theory. {\it{Rev. Mod. Phys.}} \textbf{73}, 515-562 (2001).










\end{thebibliography}

\clearpage

\bibliographystyle{NatComm.bst}

\newpage


\makeatletter
\renewcommand{\section}{\@startsection{section}{1}{0mm}
  {-\baselineskip}{0.0\baselineskip}{\bf\leftline}}
\renewcommand\thefigure{\arabic{figure}}
\renewcommand\thetable{\arabic{table}}

\renewcommand{\theequation}{S\arabic{equation}}
\renewcommand{\thefigure}{S\arabic{figure}}
\renewcommand{\thetable}{S\arabic{figure}}
\renewcommand{\thesection}{S\arabic{figure}}

\renewcommand{\bibnumfmt}[1]{[S#1]}

\setcounter{figure}{0}
\setcounter{table}{0}
\setcounter{section}{0}

\begin{center}
\large{Supplementary Information for ``Chemical bonding and Born charge in  1T-HfS$_2$''}
\end{center}

\author{S. N. Neal}
\affiliation{Department of Chemistry, University of Tennessee, Knoxville, Tennessee 37996, USA}

\author{S. Li}
\affiliation{Department of Chemical Engineering and Materials Science, University of Minnesota, Minneapolis, Minnesota, USA}

\author{T. Birol}
\affiliation{Department of Chemical Engineering and Materials Science, University of Minnesota, Minneapolis, Minnesota, USA}

\author{Janice L. Musfeldt}
\affiliation{Department of Chemistry, University of Tennessee, Knoxville, Tennessee 37996, USA}
\affiliation{Department of Physics and Astronomy, University of Tennessee, Knoxville, Tennessee 37996, USA}

\date{\today}

{
\let\clearpage\relax
\maketitle
}

\section{Supplementary discussion}

Supplementary Fig.~1 displays a close-up view of both the infrared (a) and Raman (b) spectra for 1$T$-HfS$_2$. Notice, the peak position of the infrared-active $E_u$ mode around 155 cm$^{-1}$ remains consistent in both shape, width, and peak position from 300 K to 8 K [Supplementary Fig.~1(a)]. The symmetric nature of the large $E_u$ phonon points to a homogeneous system, and therefore a high quality crystal. Supplementary Fig. 1(b) shows the slight red shift ($\approx$ 1 cm$^{-1}$) of the $A_{1g}$ mode around 340 cm$^{-1}$ as temperature decreases from 300 K to 8 K. The other Raman-active features shown ($E_g$ and $E_u$) also show a slight redshift to base temperature, with the peaks sharpening at the lowest temperatures.

\begin{figure}[tbh]
\begin{minipage}[tbh]{1.8in}
\caption{Close-up view of the (a) infrared spectra of 1T-HfS$_2$ at room temperature as well as based temperature and (b) the Raman scattering spectra from 8  to 300 K.  } \label{IR_Raman_closeup}
\end{minipage}
\begin{minipage}[tbh]{4.6in}
\includegraphics[width = 4.6in]{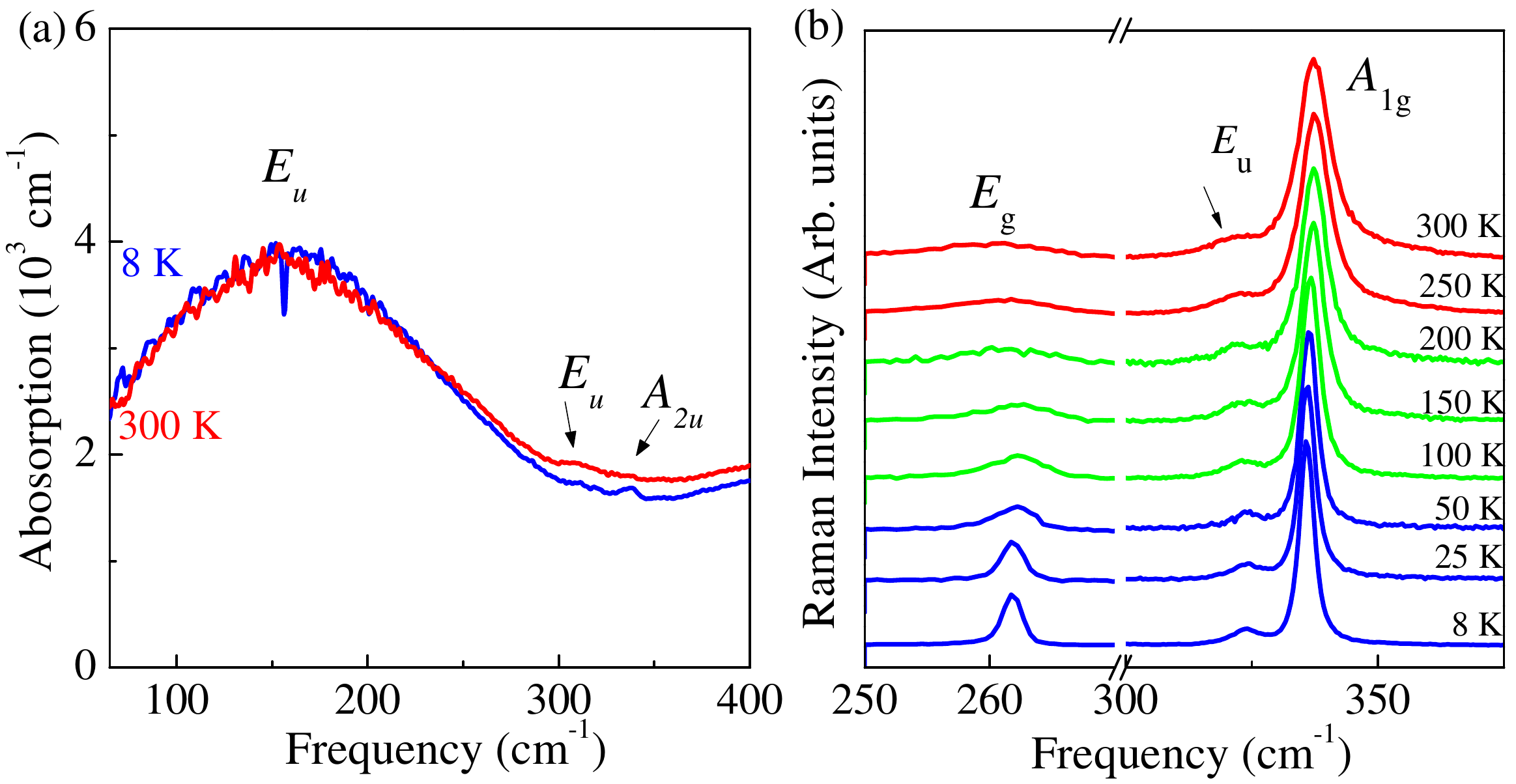}
\end{minipage}
\end{figure}
\clearpage

Supplementary Table 1 displays the FWHM values, as well as the calculated phonon lifetime values of 1T-HfS$_2$ (8-300 K) for both fundamental Raman-active phonons ($E_g$, $A_g$) discussed in the text. Recall that phonon lifetime is calculated as $  \tau = \frac{\hbar}{\Gamma}$,
where $\Gamma$ is the full width at half maximum and $\hbar$ is the reduced Planck's constant \cite{Wooten1972}. Supplementary Fig. 2 shows a general schematic detailing the relationship between FWHM ($\Gamma$), frequency ($\omega$), and phonon lifetime ($\tau$).

\begin{table}[tbh]
\caption{FWHM and phonon lifetime values for the fundamental Raman-active phonons in 1T-HfS$_2$.}

\begin{tabular}{|c||c | c || c | c ||}

\hline

&\multicolumn{2}{|c||}{\textbf{ \textit{E$_g$}}} & \multicolumn{2}{c|}{ \textbf{\textit{A$_{1g}$}}} \\ 
\hline

Temperature (K) & FWHM ($\Gamma$) (cm$^{-1}$) &  Lifetime ($\tau$) (ps)  &   FWHM ($\Gamma$) (cm$^{-1}$) & Lifetime ($\tau$) (ps) \\
\hline\hline

 8 & 260.90, 262.68 & 2.98 & 334.20, 337.64 & 1.54\\
\hline
 25 & 260.74, 262.92 & 2.44 & 333.96, 338.10 & 1.28 \\
\hline
50 & 259.55, 264.29 & 1.12  & 334.33, 338.47 & 1.28\\
\hline
100 & 259.17, 265.37 & 0.86 & 334.35, 338.98 & 1.15\\
\hline
150& 258.58, 266.39 & 0.68 & 334.58, 339.82 & 1.01 \\
\hline
 200 & 256.08, 266.37 & 0.52 & 333.44, 341.02 & 0.70 \\
 \hline
 250 & 255.17, 266.19 & 0.48 & 333.42, 341.10 & 0.69 
\\
\hline
 300 & 252.48, 266.46 & 0.38 & 333.37, 341.77 & 0.63
\\
\hline

\hline
\hline

\end{tabular}
\label{Phononlifetime}
\end{table}

\begin{figure}[tbh]
 \label{lifetime}
 \begin{minipage}{2.0 in}

\includegraphics[width = 1.5in]{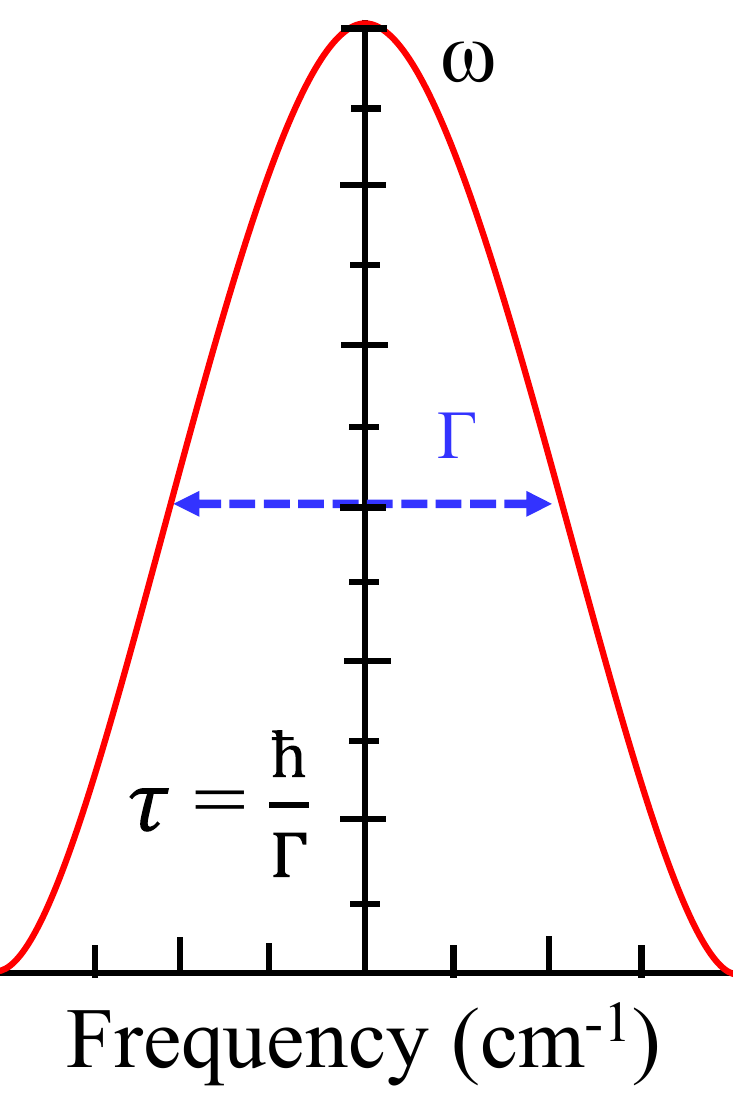}
 \end{minipage}
\begin{minipage}{3.75 in}
\caption{Displays a general schematic showing the relationship between $\Gamma$ (FWHM), $\omega$ (frequency), and $\tau$ (phonon lifetime). }
\end{minipage}
\end{figure}

\section{Supplementary methods}

Supplementary Table 2 lists the first-principles calculation results for the dielectric properties of 1T-HfS$_2$ obtained via methods outlined in the main text. By comparing the first 3 columns and second 3 columns, we can see that for the same functional and structures, the dielectric constant and Born effective charges are barely influenced by the method employed to calculate the response to external electrical field. More importantly, although the interlayer interactions are hard to reproduce accurately from first-principles due to issues with the van der Waals interaction, the DFT-relaxed structure does not affect the Born charge or susceptibility  results significantly. The effect of spin-orbit coupling is also small enough to be neglected, except on the bandgap.

Despite the similarities in the results for PBEsol, the hybrid functional HSE06 gives out a very different prediction of band gap which is much closer to the experimental one. The dielectric constant from HSE is also very different from the other ones, likely due to the larger bandgap. 

\begin{table*}
\hspace{-3cm}\begin{minipage}[b]{\linewidth}
\caption{The DFT calculation results of Born effective charge and dielectric constant.}
\label{tab:results}
\begin{tabular}{|c|p{1.5cm}|p{2cm}|p{1.5cm}|p{1.5cm}|p{2cm}|p{1.5cm}|p{1.5cm}|p{1.5cm}|p{2cm}|}
\hline
~ & PBEsol & PBEsol (structure relaxed) & PBEsol (with spin-orbital coupling & PBEsol & PBEsol (structure relaxed) & PBEsol (with SOC and structure relaxed) & meta-GGA & DFT-D3 & \textbf{HSE06} \\ \hline
Method                     & Density functional perturbation theory (DFPT)   & DFPT                            & DFPT             & Finite electric field(FE) & FE & FE & FE & FE & \textbf{FE} \\ \hline
Band gap (eV)& 0.95 & 0.95 & 1.59 & 0.95 & 0.95 & 1.53 & 1.05 & 0.95 & \textbf{2.05} \\ \hline
$\varepsilon_{xx}(\infty)$ & 10.67   & 10.91                           & 10.83    & 10.53 & 10.90 & 11.07                         & 9.94                  &10.54 &    \textbf{8.09}                   \\ \hline
$\varepsilon_{zz}(\infty)$ & 5.38   & 5.35                          & 5.40    & 5.37 & 5.35 & 5.38                         & 5.16                  &5.38 &   \textbf{4.61}                   \\ \hline
$Z_{Hf, xx}$                & 6.79   & 6.82                            & 6.85       & 6.77 & 6.82 & 6.87                       & 6.71               &6.77   &   \textbf{6.38}              \\ \hline     
$Z_{Hf, zz}$                & 1.88   & 1.87                           & 1.89       & 1.88 & 1.87 & 1.87                       & 1.92                  &1.88&   \textbf{1.96}              \\ \hline     
\end{tabular}
\end{minipage}
\end{table*}

\begin{figure}[tbh]
\begin{minipage}[tbh]{2.4in}
\caption{Comparison of the calculated density of states of (a) 1T-HfS\textsubscript{2} and (b) 1T-MoS\textsubscript{2} (c) 2H-MoS\textsubscript{2}.   } \label{fig:DOS}
\end{minipage}
\begin{minipage}[tbh]{4.0in}
\psfig{file=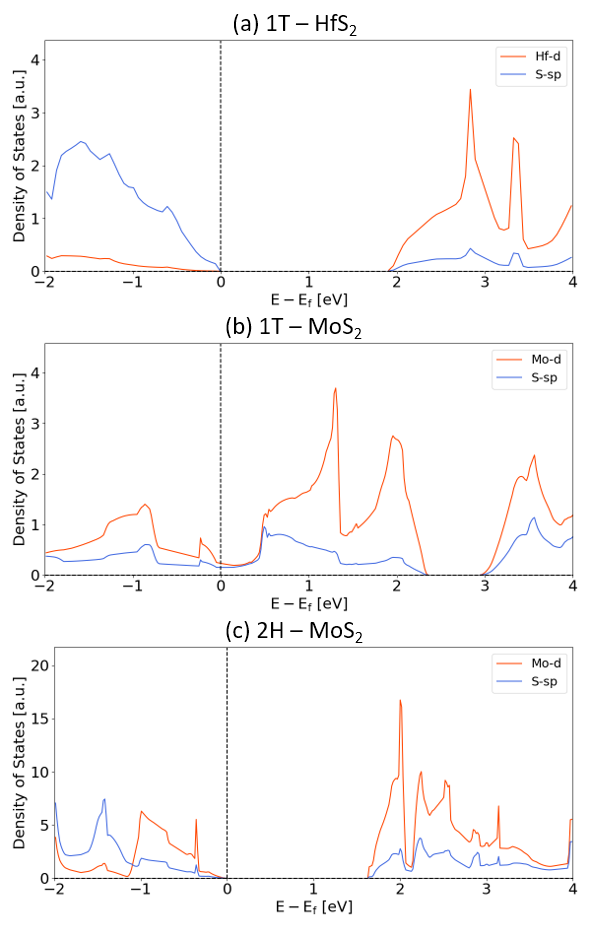,width=4.0in}
\end{minipage}
\end{figure}

The projected density states of HfS$_2$ is very different from that of MoS$_2$, because MoS$_2$ has a partially-filled Mo-4$d$ orbital. In both 1T-MoS$_2$ [metallic, Supplementary Fig. 3(b)] and 2H-MoS$_2$ [insulating, Supplementary Fig. 3(c)], bands near the Fermi level consist mostly of cation states. The mixing of both the valence and conduction band Mo-$d$ states with S-${sp}$ orbitals is nevertheless significant. This is unlike HfS$_2$ [Supplementary Fig. 3(a)], where the valence band is mostly of S character, and mixing between Hf and S states is smaller.

\begin{figure}[tbh]
\begin{minipage}[tbh]{3.4in}
\caption{One of the $E_u$ optical normal modes. Blue spheres represent the Hf atoms, and the yellow spheres represent the S atoms. The arrows are not scaled to actual value.   } \label{fig:optical}
\end{minipage}
\begin{minipage}[tbh]{3.0in}
\psfig{file=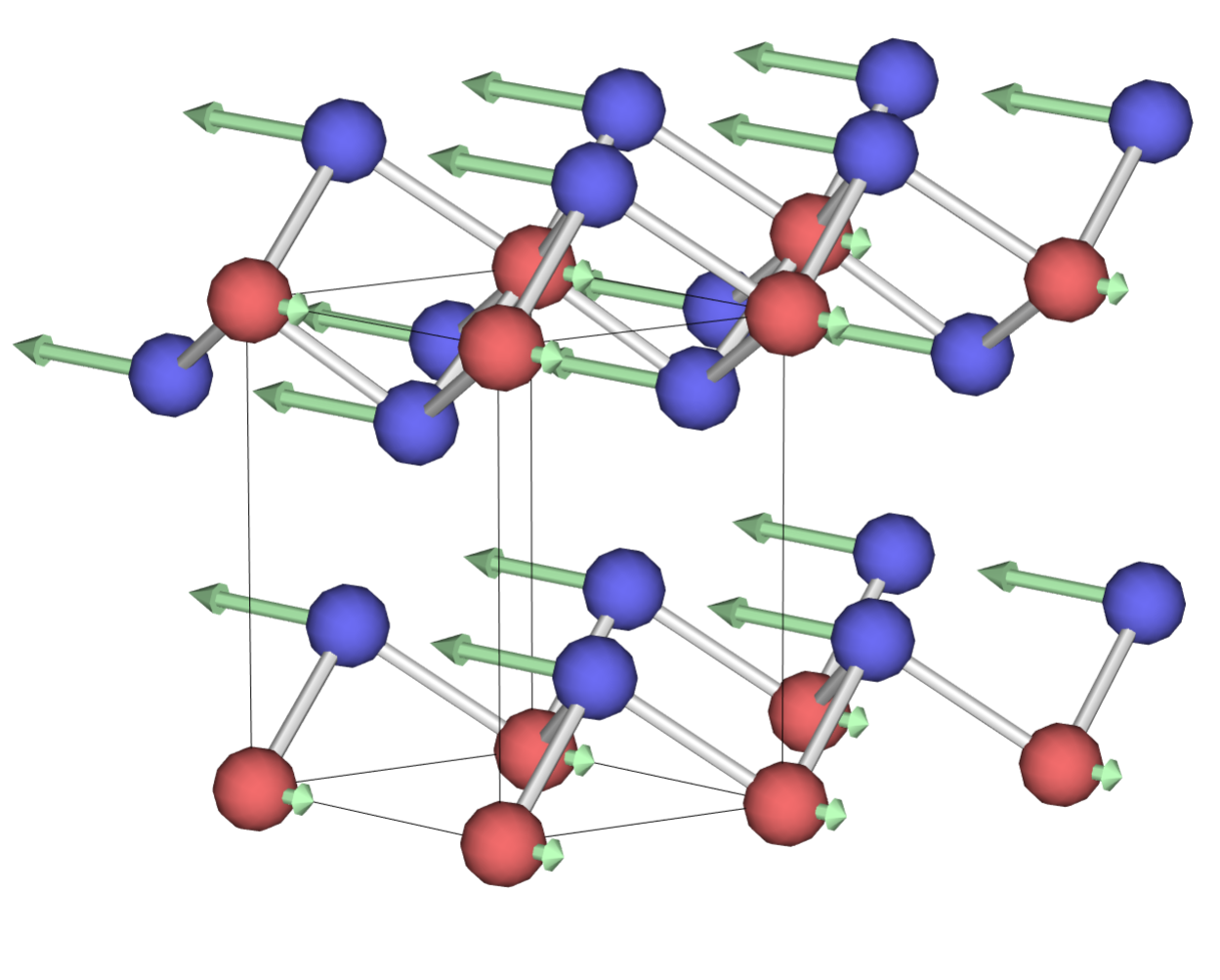,width=3.0in}
\end{minipage}
\end{figure}

The experimentally determined space group of HfS$_2$ is $P\bar{3}m1$ (\#164), which is in the trigonal crystal system. There are 2 $E_u$ phonon modes in total, of which one is optical and the other one is acoustic. The optical mode is split into two due to the LO-TO splitting. 
The form of the dynamical matrix eigenvectors for both the acoustic and the optical $E_u$ mode is set by symmetry to be: 
\begin{equation}
\tiny
\begin{blockarray}{cccccccccc}
~ & \text{Hf}_x & \text{Hf}_y & \text{Hf}_z & \text{S}_{1x} & \text{S}_{1y} & \text{S}_{1z} & \text{S}_{2x} & \text{S}_{2y} & \text{S}_{2z}\\
\begin{block}{c[ccccccccc]}
Eu_{A1} & \sqrt{m_\text{Hf}} & 0 & 0 & \sqrt{m_\text{S}} & 0 & 0 & \sqrt{m_\text{S}} & 0 & 0\\
Eu_{A2} & 0 & \sqrt{m_\text{Hf}} & 0 & 0 & \sqrt{m_\text{S}} & 0 & 0 & \sqrt{m_\text{S}} & 0\\
Eu_{O1} & \sqrt{m_\text{S}} & 0 & 0 & -\frac{1}{2}\sqrt{m_\text{Hf}} & 0 & 0 & -\frac{1}{2}\sqrt{m_\text{Hf}} & 0 & 0\\
Eu_{O2} & 0 & \sqrt{m_\text{S}} & 0 & 0 & -\frac{1}{2}\sqrt{m_\text{Hf}} & 0 & 0 & -\frac{1}{2}\sqrt{m_\text{Hf}} &0 \\
\end{block}	
\label{eq:Eu}
\end{blockarray}
\end{equation}
such that $Eu_A$ and $Eu_O$ denote acoustic and optical $E_u$ modes, respectively. 
Figure~\ref{fig:optical} shows a schematic of the optical mode. Normalizing the eigenvectors gives:
\begin{equation}
\tiny
\begin{blockarray}{cccccccccc}
~ & \text{Hf}_x & \text{Hf}_y & \text{Hf}_z & \text{S}_{1x} & \text{S}_{1y} & \text{S}_{1z} & \text{S}_{2x} & \text{S}_{2y} & \text{S}_{2z}\\
\begin{block}{c[ccccccccc]}
Eu_1 & \sqrt{\frac{2m_\text{S}}{2m_\text{S}+m_\text{Hf}}} & 0 & 0 & -\sqrt{\frac{m_\text{Hf}}{2(2m_\text{S}+m_\text{Hf})}} & 0 & 0 & -\sqrt{\frac{m_\text{Hf}}{2(2m_\text{S}+m_\text{Hf})}} & 0 & 0\\
Eu_2 & 0 & \sqrt{\frac{2m_\text{S}}{2m_\text{S}+m_\text{Hf}}} & 0 & 0 & -\sqrt{\frac{m_\text{Hf}}{2(2m_\text{S}+2m_\text{Hf})}} & 0 & 0 & -\sqrt{\frac{m_\text{Hf}}{2(2m_\text{S}+2m_\text{Hf})}} &0 \\
\end{block}
\end{blockarray}
\end{equation}
Writing the dynamical matrix on the basis of the two components of the $E_u$ mode results in a diagonal $2\times 2$ matrix with the dynamical matrix eigenvalues $D_{Eu}^{an}$ on the diagonals. 
This is the analytical contribution to the dynamical matrix, which gives the TO frequencies. In order to obtain the LO frequencies one needs to add the so-called nonanalytical contribution to the dynamical matrix as well  \cite{baroni2001phonons}:
\begin{equation}
	D^{nan}_{st, \alpha\beta}(q\to 0) = \frac{4\pi}{\Omega}e^2\mathbf{\frac{(q\cdot Z_S)_\alpha(q\cdot Z_t)_\beta}{q\cdot\bm{\varepsilon}(\infty)\cdot q}},
\end{equation}
where the $s, t$ are atomic indices and $\alpha, \beta$ are directional indices. Here, Gaussian units are used. The limit is taken as the wavevector $\mathbf{q}$ approaches zero from the direction of the polarization induced by the LO mode. 
The splitting between the transverse and longitudinal modes along a crystal axis, for example the [100] direction, can be evaluated in terms of the normalized $E_u$ dynamical matrix eigenvectors as
\begin{equation}
\omega_{TO}^2 - \omega_{LO}^2 = D_{Eu1}^{nan} = \frac{4\pi}{\Omega \varepsilon_{xx}(\infty)}e^2 \gamma_m^2.
\label{eq:LO-TO}
\end{equation}
Here $\gamma_m$ is the polarization induced by a unit displacement by the dynamical matrix eigenvalue $|\vec{u}|=1$. In this sense it is like a mode effective charge, however due to normalization of the dynamical matrix eigenvector (not the displacement) it has different units than $Z$:
\begin{equation}
	\gamma_m = \sum_i \frac{\mathbf{u_i\cdot Z_i}}{\sqrt{m_i}},
\end{equation}
where $u_i$ is the $i^{th}$ component of the dynamical matrix eigenvector. In this case, the eigenvector is the normalized one for the $E_u$ mode, which is shown in the first row of Equation~\ref{eq:Eu}.

In the specific (diatomic) case of HFS$_2$, we have $Z_{\text{Hf},xx}=-2Z_{\text{S},xx}$, thus the effective mode would be:
\begin{equation}
\begin{aligned}
	\gamma_m &= \sqrt{\frac{2m_\text{S}}{2m_\text{S}+m_\text{Hf}}}\frac{Z_{\text{Hf},xx}}{\sqrt{m_\text{Hf}}} -  2\sqrt{\frac{m_\text{Hf}}{2(2m_\text{S}+m_\text{Hf})}}\frac{Z_{\text{S}, xx}}{\sqrt{m_\text{S}}} \\
	&=  \left(\sqrt{\frac{2m_\text{S}}{2m_\text{S}+m_\text{Hf}}}\frac{1}{\sqrt{m_\text{Hf}}} +  \sqrt{\frac{m_\text{Hf}}{2(2m_\text{S}+m_\text{Hf})}}\frac{1}{\sqrt{m_\text{S}}}\right)Z_{\text{Hf},xx} \\
	&=\frac{1}{\sqrt{m^*}}Z_{\text{Hf},xx}
\end{aligned}
\end{equation}
Here the $m^*$ is the effective mass, which has the relationship: 
\begin{equation}
    \frac{1}{m^*} = \left(\sqrt{\frac{2m_\text{S}}{2m_\text{S}+m_\text{Hf}}}\frac{1}{\sqrt{m_\text{Hf}}} +  \sqrt{\frac{m_\text{Hf}}{2(2m_\text{S}+m_\text{Hf})}}\frac{1}{\sqrt{m_\text{S}}}\right)^2 = \frac{1}{m_\text{Hf}}+\frac{1}{2m_\text{S}}
\end{equation}
which gives $47.02~u$ for 1T-HfS$_2$.

In the end, the Born effective charge can be simplified using this effective mass from Equation~\ref{eq:LO-TO}:
\begin{equation}
\omega_{TO}^2 - \omega_{LO}^2 = \frac{4\pi}{\Omega \varepsilon_{xx}(\infty)}e^2 \frac{Z_{\text{Hf},xx}^2}{m^*}
\end{equation}

HfS$_2$ is an ionic crystal with a relatively simple band structure: the conduction band is mainly of Hf-$d$ character, and the valence band is mainly of S-$p$ character. This simplifies the Wannierization process as the bands of interest are not entangled with each other. As shown in Supplementary Fig. 5, the band structure reconstructed from the Wannier based tight binding model matches with the DFT one perfectly (except at very high energies). 

\begin{figure}[h!]
\begin{minipage}[tbh]{2.4in}
\caption{The band structure calculated from DFT and interpolated from Wannier functions. PBEsol is used here. The dark blue lines are directly from a DFT-calculated results. The thicker lines above are interpolated from Wannier functions.} \label{fig:wannier}
\end{minipage}
\begin{minipage}[tbh]{4.0in}
\psfig{file=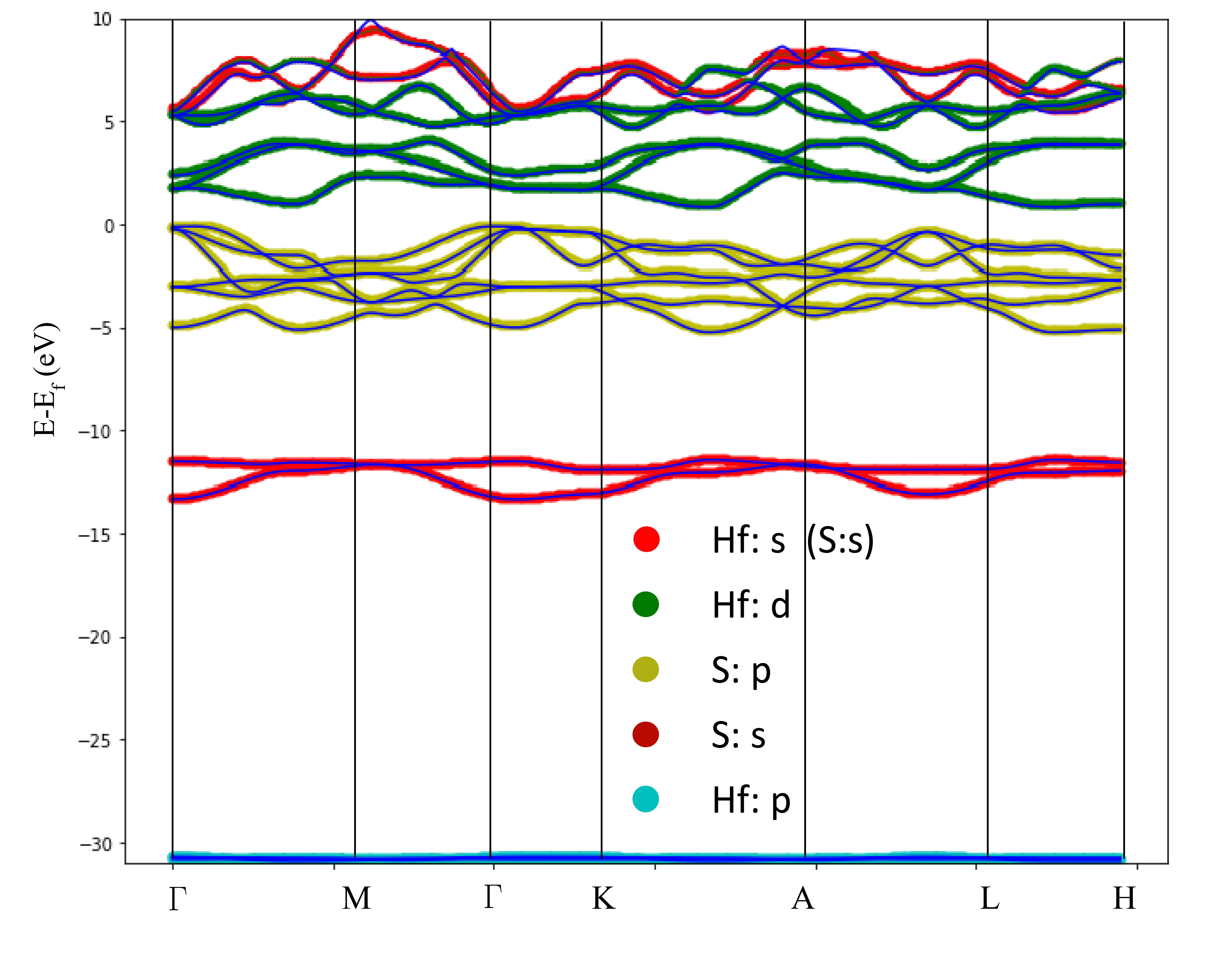,width=4.0in}
\end{minipage}
\end{figure}

In the main text, only one S-$p$ orbital (oriented along the z-direction) is shown. The other two S-$p$ orbitals are  similar to the $p_z$ orbital in the sense that they also hybridize with the nearest neighbor Hf cation [Supplementary Fig. 6]. The 3 other $s$-$p$ orbitals centered on the other S anion are not shown here, since they are symmetry equivalent to those in Supplementary Fig. 6.

\begin{figure}[tbh]
\begin{minipage}[tbh]{2.2in}
\caption{The Wannier functions of the $s$-$p$ orbitals in HfS$_2$. The colors represent the sign of the Wannier functions.} \label{fig:pxyz}
\end{minipage}
\begin{minipage}[tbh]{4.2in}
\psfig{file=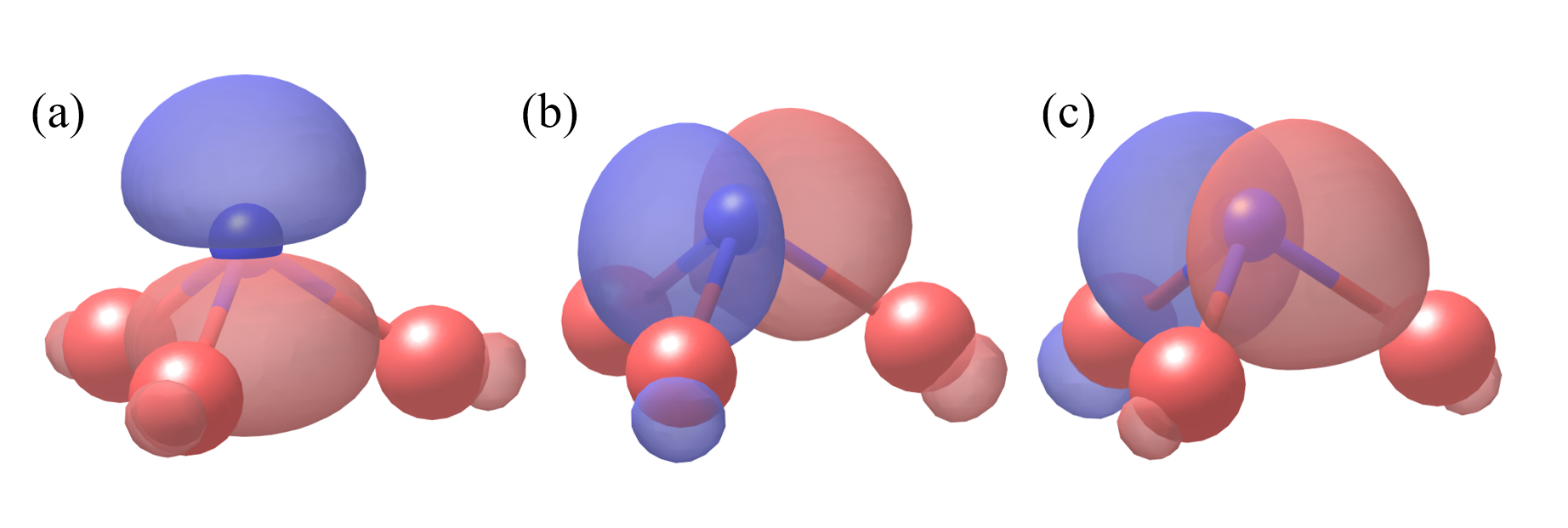,width=4.2in}
\end{minipage}
\end{figure}
\clearpage

\bibliographystyle{Nature}

\end{document}